\documentclass[prd,twocolumn,preprintnumbers,superscriptaddress,nofootinbib]{revtex4}

\usepackage [english]{babel}
\usepackage [autostyle, english = american]{csquotes}
\MakeOuterQuote{"}
\usepackage{tabularx} 

\usepackage[title]{appendix}
\usepackage{graphicx} 
\usepackage{dcolumn} 
\usepackage{bm} 
\usepackage{amssymb,amsmath} 
\usepackage{multirow}
\usepackage{lineno}
\usepackage{pdfpages}
\usepackage{capt-of}
\usepackage{float}
\usepackage{heppennames}
\usepackage[hyperfootnotes=false]{hyperref}

\usepackage{placeins}

\begin{document}


\newcommand{\beq}{\begin{equation}}
\newcommand{\eeq}{\end{equation}}
\newcommand{\D}{\displaystyle}
\newcommand{\panelwidth}{0.99\columnwidth}
\newcommand{\nuwro}{\textsc{NuWro}}
\newcommand{\achilles} {{\sc{achilles}}}

\newcommand{\minerva}{MINER$\nu$A}
\newcommand{\ttbs}{\char'134}
\newcommand{\integ}{\iint}
\newcommand{\FA}{${\cal F}_A$}
\newcommand{\fa}{${\cal F}_A(q^2)$}
\newcommand{\gepQ}{$G_{Ep}(q^2)$}
\newcommand{\nubar}[0]{$\overline{\nu}$}

\newcommand{\gr}[1]{\textcolor{lightgray}{#1}}

\newcommand{\repeatcaption}[2]{%
  \captionsetup{list=no}%
  \caption{#2 (repeated from page \pageref{#1})}%
}

\newcommand{\EqInSeq}[1]{\texorpdfstring{$#1$}{Lg}}

\newcommand{\azero} {\ensuremath{A_0}\xspace}
\newcommand{\atwo} {\ensuremath{A_2}\xspace}
\newcommand{\thetaCS} {\ensuremath{\theta_\mathrm{CS}}\xspace}
\newcommand{\qqbar}{\ensuremath{\Pq\Paq}\xspace}
\newcommand{\gluglu}{\ensuremath{\Pg\Pg}\xspace}
\newcommand{\qglu}{\ensuremath{\Pq\Pg}\xspace}
\newcommand{\DY}{\ensuremath{\gamma^*/\PZ}\xspace}
\newcommand{\azeroqq}{\ensuremath{A_0^{\Pq\bar{\Pq}}}\xspace}
\newcommand{\azerogq}{\ensuremath{A_0^{\Pg\Pq}}\xspace}
\newcommand{\atwoqq}{\ensuremath{A_2^{\Pq\bar{\Pq}}}\xspace}
\newcommand{\atwogq}{\ensuremath{A_2^{\Pg\Pq}}\xspace}

\newcommand{\mll}{\ensuremath{m_{\ell\ell}}\xspace}
\newcommand{\Mmumu}{\ensuremath{M_{\Pgm\Pgm}}\xspace}
\newcommand{\yll}{\ensuremath{y_{\ell\ell}}\xspace}
\newcommand{\ptll}{\ensuremath{\pt^{\ell\ell}}\xspace}
\newcommand{\yZ}{\ensuremath{y_{\PZ}}\xspace}
\newcommand{\ptZ}{\ensuremath{p_{\text{T}}^{\PZ}}\xspace}
\newcommand{\pthat}{\ensuremath{\hat{p}_{\text{T}}}\xspace}
\newcommand{\rthat}{\ensuremath{\hat{R}_{\text{T}}}\xspace}
\newcommand{\ptsq}{\ensuremath{p_{\text{T}}^2}\xspace}
\newcommand{\ptjet}{\ensuremath{p_{\text{T}}^{\text{jet}}}\xspace}
\newcommand{\vecpt}{\ensuremath{\vec{p}_{\text{T}}}\xspace}
\newcommand{\qt}{\ensuremath{q_{\text{T}}}\xspace}


\newcommand{\AFB}{\ensuremath{A_\text{FB}}\xspace}

\newcommand{\zll} {\ensuremath{\PZ/\gamma^*\to\Pl\Pl}\xspace}
\newcommand{\uu} {\ensuremath{\Pgm\Pgm}\xspace}
\newcommand{\ue} {\ensuremath{\Pgm\Pe}\xspace}
\newcommand{\eu} {\ensuremath{\Pe\Pgm}\xspace}
\newcommand{\ee} {\ensuremath{\Pe\Pe}\xspace}
\newcommand{\eg} {\ensuremath{\Pe\Pg}\xspace}
\newcommand{\eh} {\ensuremath{\Pe\Ph}\xspace}
\newcommand{\ug} {\ensuremath{\Pgm\Pg}\xspace}
\newcommand{\uh} {\ensuremath{\Pgm\Ph}\xspace}
\newcommand{\ztt} {\ensuremath{\PZ/\gamma\to\tautau}\xspace}

\newcommand{\nvtx}{\ensuremath{N_{vtx}}\xspace}
\newcommand{\meh}{\ensuremath{M_{eh}}\xspace}
\newcommand{\electronEndcapEta}{\ensuremath{1.57<|\eta|<2.50}\xspace}
\newcommand{\hele}{\ensuremath{h}\xspace}

\newcommand{\gele}{\ensuremath{g}\xspace}
\newcommand{\ptg}{\ensuremath{p_{\mathrm{T}}^{\gamma}}\xspace}
\newcommand{\sieie}{\ensuremath{\sigma_{i\eta i\eta}}\xspace}
\newcommand{\higheta}{high-$\eta$\xspace}
\newcommand{\recophoton}{$\gamma_R$\xspace}
\newcommand{\genphoton}{$\gamma_G$\xspace}
\newcommand{\meg}{\ensuremath{M_{eg}}\xspace}

\newcommand{\Iu} {\ensuremath{\tilde{\Pgm}}\xspace}
\newcommand{\Ie} {\ensuremath{\tilde{\Pe}}\xspace}
\newcommand{\Ig} {\ensuremath{\tilde{\gele}}\xspace}
\newcommand{\Ih} {\ensuremath{\tilde{\hele}}\xspace}

\newcommand{\ETm}{\ensuremath{E_{\mathrm{T}}^{\text{miss}}}\xspace}
\newcommand{\MET}{\ETm}
\newcommand{\ETmiss}{\ETm}

\newcommand{\jets}{\ensuremath{\text{jets}}}
\newcommand{\ttNew}{\ensuremath{\mathrm{t}\bar{\mathrm{t}}}\xspace}
\newcommand{\ttbar}{\ensuremath{\mathrm{t\bar{t}}}\xspace}
\newcommand{\ttb}{\ensuremath{\mathrm{t\bar{t}}}\xspace}
\newcommand{\ptmiss}{\ensuremath{{p}_{\text T}^\text{miss}}\xspace}
\newcommand{\ptmissv}{\ensuremath{\boldsymbol{p}_{\text T}^\text{miss}}\xspace}
\newcommand{\nuv}{\ensuremath{\boldsymbol{\nu}}}

\newcommand{\ppbar}{\ensuremath{\mathrm{p}\bar{\mathrm{p}}}\xspace}
\newcommand{\pp}{\ensuremath{\mathrm{p}\mathrm{p}}\xspace}
\newcommand{\tW}{\ensuremath{\mathrm{tW}}}
\newcommand{\PQb}{\ensuremath{\mathrm{b}}}
\newcommand{\PQt}{\ensuremath{\mathrm{t}}}
\newcommand{\gj}{\ensuremath{\gamma + \jets}\xspace}
\newcommand{\zj}{\ensuremath{\mathrm{Z} + \jets}\xspace}

\newcommand{\delY}{\ensuremath{|y_{\mathrm{t}}-y_{\bar{\mathrm{t}}}|}}
\newcommand{\Pt}            {\ensuremath{{p_{\text T}}}\xspace}
\newcommand{\pt}            {\ensuremath{{p_{\text T}}}\xspace}
\newcommand{\PtVec}         {\ensuremath{{\vec{p}_{\text T}}}\xspace}
\newcommand{\PVec}          {\ensuremath{{\vec{p}}}\xspace}
\newcommand{\etaabs}        {\ensuremath{|\eta|}}
\newcommand{\sqrts}        {\ensuremath{\sqrt{s}}\xspace}
\newcommand{\metVec}       {\ensuremath{\slashed{\vec{E}}_{T}}\xspace}
\newcommand{\ptvecmiss}		{\metVec}
\newcommand{\Etji}[1]       {\ensuremath{\Et^{\ji{#1}}}\xspace}
\newcommand{\Ptji}[1]       {\ensuremath{\Pt^{\ji{#1}}}\xspace}
\newcommand{\PtVecji}[1]    {\ensuremath{\PtVec^{\ji{#1}}}\xspace}

\newcommand{\ji}[1]         {\ensuremath{\rm j_{#1}}\xspace}

\newcommand\rs{\raisebox{1.0ex}[-1.0ex]}
\newcommand{\ra}{\ensuremath{\rightarrow}}
\newcommand{\mur}{\ensuremath{\mu_\mathrm{r}}\xspace}
\newcommand{\muf}{\ensuremath{\mu_\mathrm{f}}\xspace}
\newcommand{\AMCATNLO}{\textsc{mg}5\_\text{a}\textsc{mc@nlo}\xspace}
\newcommand{\OPENLOOPS}{\textsc{OpenLoops}\xspace}

\newcommand{\Mbl}{\ensuremath{M_\mathrm{b\ell}}\xspace}
\newcommand{\yt}{\ensuremath{Y_\mathrm{t}}\xspace}
\newcommand{\tq}{\ensuremath{\mathrm{t}}\xspace}
\newcommand{\bq}{\ensuremath{\mathrm{b}}\xspace}
\newcommand{\Wb}{\ensuremath{\mathrm{W}}\xspace}

\newcommand{\DYbl}{\ensuremath{\Delta y_\mathrm{b\ell}}\xspace}
\newcommand{\Dybl}{\ensuremath{\Delta y_\mathrm{b\ell}}\xspace}

\newcommand{\absDybl}{\ensuremath{|\Delta y|_\mathrm{b\ell}}\xspace}

\newcommand{\xfitter}{\ensuremath{\textsc{xFitter}}\xspace}

\newenvironment{scotch}[1]{\protect\centering\tabular{#1}\hline\hline}{\hline\endtabular}


\newcommand{\pbinv} {\mbox{\ensuremath{\,\text{pb}^\text{$-$1}}}\xspace}
\newcommand{\fbinv} {\mbox{\ensuremath{\,\text{fb}^\text{$-$1}}}\xspace}

\newcommand{\de}{\ensuremath{^\circ}}
\newcommand{\ten}[1]{\ensuremath{\times \text{10}^\text{#1}}}
\newcommand{\unit}[1]{\ensuremath{\text{\,#1}}\xspace}
\newcommand{\mum}{\ensuremath{\,\mu\text{m}}\xspace}
\newcommand{\micron}{\ensuremath{\,\mu\text{m}}\xspace}
\newcommand{\cm}{\ensuremath{\,\text{cm}}\xspace}
\newcommand{\mm}{\ensuremath{\,\text{mm}}\xspace}
\newcommand{\mus}{\ensuremath{\,\mu\text{s}}\xspace}
\newcommand{\keV}{\ensuremath{\,\text{ke\hspace{-.08em}V}}\xspace}
\newcommand{\MeV}{\ensuremath{\,\text{Me\hspace{-.08em}V}}\xspace}
\newcommand{\MeVns}{\ensuremath{\text{Me\hspace{-.08em}V}}\xspace} 
\newcommand{\GeV}{\ensuremath{\,\text{Ge\hspace{-.08em}V}}\xspace}
\newcommand{\GeVns}{\ensuremath{\text{Ge\hspace{-.08em}V}}\xspace} 
\newcommand{\gev}{\GeV}
\newcommand{\TeV}{\ensuremath{\,\text{Te\hspace{-.08em}V}}\xspace}
\newcommand{\TeVns}{\ensuremath{\text{Te\hspace{-.08em}V}}\xspace} 
\newcommand{\PeV}{\ensuremath{\,\text{Pe\hspace{-.08em}V}}\xspace}
\newcommand{\keVc}{\ensuremath{{\,\text{ke\hspace{-.08em}V\hspace{-0.16em}/\hspace{-0.08em}}c}}\xspace}
\newcommand{\MeVc}{\ensuremath{{\,\text{Me\hspace{-.08em}V\hspace{-0.16em}/\hspace{-0.08em}}c}}\xspace}
\newcommand{\GeVc}{\ensuremath{{\,\text{Ge\hspace{-.08em}V\hspace{-0.16em}/\hspace{-0.08em}}c}}\xspace}
\newcommand{\GeVcns}{\ensuremath{{\text{Ge\hspace{-.08em}V\hspace{-0.16em}/\hspace{-0.08em}}c}}\xspace} 
\newcommand{\TeVc}{\ensuremath{{\,\text{Te\hspace{-.08em}V\hspace{-0.16em}/\hspace{-0.08em}}c}}\xspace}
\newcommand{\keVcc}{\ensuremath{{\,\text{ke\hspace{-.08em}V\hspace{-0.16em}/\hspace{-0.08em}}c^\text{2}}}\xspace}
\newcommand{\MeVcc}{\ensuremath{{\,\text{Me\hspace{-.08em}V\hspace{-0.16em}/\hspace{-0.08em}}c^\text{2}}}\xspace}
\newcommand{\GeVcc}{\ensuremath{{\,\text{Ge\hspace{-.08em}V\hspace{-0.16em}/\hspace{-0.08em}}c^\text{2}}}\xspace}
\newcommand{\GeVccns}{\ensuremath{{\text{Ge\hspace{-.08em}V\hspace{-0.16em}/\hspace{-0.08em}}c^\text{2}}}\xspace} 
\newcommand{\TeVcc}{\ensuremath{{\,\text{Te\hspace{-.08em}V\hspace{-0.16em}/\hspace{-0.08em}}c^\text{2}}}\xspace}

\newcommand{\pb}{\ensuremath{pb^{-1}}}
\newcommand{\tev}{\TeV}

\newcommand{\kt}		{\ensuremath{\mathrm{k}_{\mathrm{T}}}}

\newcommand {\etal}{\mbox{et al.}\xspace} 
\newcommand {\ie}{\mbox{i.e.}\xspace}     
\newcommand {\etc}{\mbox{etc.}\xspace}     
\newcommand {\vs}{\mbox{\textsl{vs.}}\xspace}      
\newcommand {\mdash}{\ensuremath{\text{---}}} 
\providecommand {\NA}{\ensuremath{\text{---}}}    

\newcommand {\Lone}{Level-1\xspace} 
\newcommand {\Ltwo}{Level-2\xspace}
\newcommand {\Lthree}{Level-3\xspace}

\newcommand{\doublet}[2]{
  \left(\!
    \begin{array}{c}
      {#1} \\
      {#2}
    \end{array}
    \!\right) 
}

\newcommand{\triplet}[3]{
  \left(\!
    \begin{array}{c}
      {#1} \\
      {#2} \\
      {#3}
    \end{array}
    \!\right) 
}

\newcommand{\ta}{\mbox{$\tau$}}
\newcommand{\qq}{$\mathrm{q}\bar{\mathrm{q}}$~}
\newcommand{\met}{\ensuremath{E_\mathrm{T}^{\mathrm{miss}}}}
\newcommand{\st}{\ensuremath{S_\mathrm{T}}}
\newcommand{\xsect}{cross section} 
\newcommand{\xSect}{Cross section} 
\newcommand{\Nttbar}{N_{\mathrm{t}\overline{\mathrm{t}}}}
\newcommand{\sigttbar}{\sigma_{\mathrm{t}\overline{\mathrm{t}}}}
\newcommand{\rec}{\mathrm{rec}}
\newcommand{\gen}{\mathrm{gen}}
\newcommand{\mt}{\ensuremath{M_T^W}}
\newcommand{\wpt}{\ensuremath{p_T^W}}

\newcommand{\W}{$\mathrm{W}$\xspace}
\newcommand{\see}{\ensuremath{\sigma_{i\eta i\eta}}}
\newcommand{\nfp}{\ensuremath{I_{\mathit{photon},\mathit{nfp}}}}
\newcommand{\CL}{\ensuremath{\mathit{CL}}}
\newcommand{\Sh}{$\sqrt{s} = 13$\,TeV\xspace}
\newcommand{\AGC}{\textit{aTGC}\xspace}
\newcommand{\TP}{\textit{T\&P}\xspace}
\newcommand{\ISR}{\textit{ISR}\xspace}
\newcommand{\FSR}{\textit{FSR}\xspace}
\newcommand{\ECAL}{\textit{ECAL}\xspace}
\newcommand{\HCAL}{\textit{HCAL}\xspace}
\newcommand{\sherpa}{\textit{Sherpa}\xspace}
\newcommand{\mcfm}{\textit{MCFM}\xspace}
\newcommand{\MC}{MC\xspace}
\newcommand{\FIG}[1]{Fig.\,\ref{#1}\xspace}
\newcommand{\FORM}[1]{Eq.\,\ref{#1}\xspace}
\newcommand{\TAB}[1]{Table\,\ref{#1}\xspace}
\newcommand{\sihih}{\ensuremath{\sigma_\mathrm{i\eta i\eta}}}
\newcommand{\alphas}{\ensuremath{\alpha_\mathrm{s}}}
\newcommand{\nback}{\ensuremath{N_{\mathrm{bkg}}}}
\newcommand{\nsig}{\ensuremath{N_{\mathrm{sig}}}}
\newcommand{\tqh}{\ensuremath{\mathrm{t}_\mathrm{h}}\xspace}
\newcommand{\tql}{\ensuremath{\mathrm{t}_\ell}\xspace}
\newcommand{\tbq}{\ensuremath{\mathrm{\bar{t}}}\xspace}
\newcommand{\Mtop}{\ensuremath{m_\mathrm{t}}\xspace}
\newcommand{\MW}{\ensuremath{m_\mathrm{W}}\xspace}
\newcommand{\lpj}{$\ell$+jets\xspace}
\newcommand{\UL}{pb$^{-1}$\xspace}
\newcommand{\fUL}{fb$^{-1}$\xspace}
\newcommand{\Dn}{\ensuremath{D_{\nu,\mathrm{min}}}\xspace}
\providecommand{\AMCATNLO}{\textsc{mg}5\_\text{a}\textsc{mc@nlo}\xspace}
\newcommand{\PYTHIAG}{\textsc{Pythia}\xspace}
\newcommand{\PYTHIAA}{\textsc{Pythia8}\xspace}
\newcommand{\GEANTF}{\textsc{Geant4}\xspace}
\newcommand{\HERWIGPP}{\textsc{Herwig++}\xspace}
\newcommand{\MPT}{\ensuremath{\vec{p}_\mathrm{T}^\mathrm{miss}}\xspace}
\newcommand{\Hathor}{\textsc{Hathor}\xspace}
\newcommand{\yukawa}{\ensuremath{Y_{t}}\xspace}
\newcommand{\deltaY}{\ensuremath{\Delta y_{t}}\xspace}
\newcommand{\absDelY}{\ensuremath{|\Delta y_{\ttb}|}\xspace}
\newcommand{\Mttb}{\ensuremath{M_{\ttb}}\xspace}
\newcommand{\Ryukawa}{\ensuremath{}}

\newcommand {\sinw} {\ensuremath{\sin^2\theta_\mathrm{W}}\xspace}
\newcommand {\sineff} {\ensuremath{\sin^2\theta_\mathrm{eff}^\ell}\xspace}
\newcommand {\sinefff} {\ensuremath{\sin^2\theta_\mathrm{eff}^\ell}\xspace}
\newcommand {\afour} {\ensuremath{A_4}\xspace}
\newcommand {\cs} {\ensuremath{\cos\theta_\mathrm{CS}}\xspace}
\newcommand {\phiCS} {\ensuremath{\phi_\mathrm{CS}}\xspace}
\renewcommand {\ll} {\ensuremath{{\ell\ell}}\xspace}
\newcommand {\tautau} {\ensuremath{{\tau\tau}}\xspace}
\newcommand {\afb} {\ensuremath{A_\mathrm{FB}}\xspace}
\newcommand {\afbw} {\ensuremath{A_\mathrm{FB}^\mathrm{w}}\xspace}
\newcommand{\abs}[1]{\ensuremath{\lvert #1 \rvert}}

\newcommand{\ACERMC} {\textsc{AcerMC}\xspace}
\newcommand{\ALPGEN} {{\textsc{alpgen}}\xspace}
\newcommand{\BLACKHAT} {{\textsc{BlackHat}}\xspace}
\newcommand{\CALCHEP} {{\textsc{CalcHEP}}\xspace}
\newcommand{\CHARYBDIS} {{\textsc{charybdis}}\xspace}
\newcommand{\CMKIN} {\textsc{cmkin}\xspace}
\newcommand{\CMSIM} {{\textsc{cmsim}}\xspace}
\newcommand{\CMSSW} {{\textsc{cmssw}}\xspace}
\newcommand{\COBRA} {{\textsc{cobra}}\xspace}
\newcommand{\COCOA} {{\textsc{cocoa}}\xspace}
\newcommand{\COMPHEP} {\textsc{CompHEP}\xspace}
\newcommand{\EVTGEN} {{\textsc{evtgen}}\xspace}
\newcommand{\FAMOS} {{\textsc{famos}}\xspace}
\newcommand{\FASTJET} {{\textsc{FastJet}}\xspace}
\newcommand{\FEWZ} {{\textsc{fewz}}\xspace}
\newcommand{\GARCON} {\textsc{garcon}\xspace}
\newcommand{\GARFIELD} {{\textsc{garfield}}\xspace}
\newcommand{\GEANE} {{\textsc{geane}}\xspace}
\newcommand{\GEANTfour} {{\textsc{Geant4}}\xspace}
\newcommand{\GEANTthree} {{\textsc{geant3}}\xspace}
\newcommand{\GEANT} {{\textsc{geant}}\xspace}
\newcommand{\HDECAY} {\textsc{hdecay}\xspace}
\newcommand{\HERWIG} {{\textsc{herwig}}\xspace}
\newcommand{\HERWIGpp} {{\textsc{herwig++}}\xspace}
\newcommand{\POWHEG} {{\textsc{powheg}}\xspace}
\newcommand{\POWHEGEW}{{\textsc{powheg-ew}}\xspace}
\newcommand{\powhegzew}{\POWHEG-\textsc{z}\_ew\xspace}
\newcommand{\DYNNLO} {{\textsc{DYNNLO}}\xspace}

\newcommand{\HIGLU} {{\textsc{higlu}}\xspace}
\newcommand{\HIJING} {{\textsc{hijing}}\xspace}
\newcommand{\HYDJET} {{\textsc{hydjet}}\xspace}
\newcommand{\IGUANA} {\textsc{iguana}\xspace}
\newcommand{\ISAJET} {{\textsc{isajet}}\xspace}
\newcommand{\ISAPYTHIA} {{\textsc{isapythia}}\xspace}
\newcommand{\ISASUGRA} {{\textsc{isasugra}}\xspace}
\newcommand{\ISASUSY} {{\textsc{isasusy}}\xspace}
\newcommand{\ISAWIG} {{\textsc{isawig}}\xspace}
\newcommand{\MADGRAPH} {\textsc{MadGraph}\xspace}
\newcommand{\MCATNLO} {\textsc{mc@nlo}\xspace}
\newcommand{\MCFM} {\textsc{mcfm}\xspace}
\newcommand{\MILLEPEDE} {{\textsc{millepede}}\xspace}
\newcommand{\ORCA} {{\textsc{orca}}\xspace}
\newcommand{\OSCAR} {{\textsc{oscar}}\xspace}
\newcommand{\PHOTOS} {\textsc{photos}\xspace}
\newcommand{\PROSPINO} {\textsc{prospino}\xspace}
\newcommand{\PYTHIA} {{\textsc{pythia}}\xspace}
\newcommand{\SHERPA} {{\textsc{sherpa}}\xspace}
\newcommand{\TAUOLA} {\textsc{tauola}\xspace}
\newcommand{\TOPREX} {\textsc{TopReX}\xspace}
\newcommand{\XDAQ} {{\textsc{xdaq}}\xspace}
\newcommand{\MGvATNLO}{\MADGRAPH{}5\_a\MCATNLO}
\newcommand{\TOPpp}{\textsc{Top++}\xspace}
\newcommand{\GENEVA}{\textsc{geneva}\xspace}
\newcommand{\minlo}{\textsc{MiNLO}\xspace}
\newcommand{\minnlo}{\textsc{MiNNLO}\xspace}
\newcommand{\minnlops}{\ensuremath{\text{\sc MiNNLO}_{\text{PS}}}\xspace}

\newcommand{\wjet} {{\ensuremath{\PW+jets}}\xspace}

\newcommand{\stat}{\ensuremath{\,\text{(stat)}}\xspace}
\newcommand{\syst}{\ensuremath{\,\text{(syst)}}\xspace}
\newcommand{\thy}{\ensuremath{\,\text{(theo)}}\xspace}
\newcommand{\pdf}{\ensuremath{\,\text{(PDF)}}\xspace}
\newcommand{\exper}{\ensuremath{\,\text{(exp)}}\xspace} 

 \newcommand{\feta} {\ensuremath{|\eta|}\xspace}

\newcommand{\Rochester}{Department of Physics and Astronomy, University of Rochester, Rochester, NY 14627, USA}
\newcommand{\Korea}{Department of Physics and Astronomy, Seoul National University, Korea}

\title{Investigation of the difference in the angular distributions of $Z\to \ell^{+}\ell^{-}$ events produced in quark-antiquark, quark-gluon and gluon-gluon collisions
}

\author{Arie~Bodek}
\affiliation{\Rochester}
\email{bodek@pas.rochester.edu}
\author{Giulia-Maria Bulugean}
\affiliation{\Rochester}
\author{Aran Garcia-Bellido}
\affiliation{\Rochester}
\author{Hyon San Seo}
\affiliation{\Rochester}
\author{Rhys Taus}
\affiliation{\Rochester}
\author{Un-Ki Yang}
\affiliation{\Korea}

\date{\today}

\begin{abstract}

Measurements of the angular distributions of \PZ boson-decay leptons at the Large Hadron Collider (LHC) are of interest in QCD studies and precision electroweak measurements. \PZ boson production at the LHC is dominated by gluon processes and the quark-antiquark (\qqbar) process accounts for only 40\% of the cross section. We investigate the theoretical predictions (using the \POWHEG-\minnlops event generator) for the difference in the angular distributions of $\pp\to \gamma^{*}/Z \to \ell^+\ell^-$ 
($ \Pgm^+\Pgm^-$ or $\Pe^{+}\Pe^{-}$)
events produced via \qqbar, quark-gluon (\qglu), and gluon-gluon (\gluglu) processes. The study is done for proton-proton collisions at the Large Hadron Collider (LHC) at $\sqrt{s} = 13\TeV$. We investigate the angular coefficients \azero and \atwo (for \thetaCS and \phiCS distributions in the Collins-Soper frame) for the different processes as a function of the \PZ boson rapidity (\yZ) transverse momentum (\ptZ) and final state jet multiplicity, and propose how these can be investigated experimentally at the LHC.

\end{abstract}


\maketitle
\section{Introduction}

Measurements of the angular distributions of leptons from electroweak vector-boson decays at the Large Hadron Collider (LHC), including both \PW and \PZ boson production, provide tests of perturbative QCD and inputs to precision electroweak measurements such as the \PW boson mass and the effective weak mixing angle~\cite{CMS:2024lrd,CMS:2024ony}. These distributions are conventionally expressed in terms of angular coefficients, which describe the polarization state of the vector boson and its dependence on production kinematics, including the boson transverse momentum generated by QCD radiation~\cite{Mirkes:1994eb,ATLASWang}.

As described below,  existing measurements of the \PZ boson angular coefficients~\cite{ATLASang,CMSang,CMS:2026amb,LHCb:2022tbc} by ATLAS,  CMS
and LHCb show that several coefficients are well described by QCD predictions, while the measured difference $A_0-A_2$, which quantifies the violation of the Lam-Tung relation~\cite{LTLam:1978zr}($A_0-A_2=0$), is larger than predicted. These discrepancies indicate that current simulations do not fully describe all features of the measured angular distributions, and suggest that a more differential understanding of the angular coefficients is needed.

\begin{figure}
\centering
\includegraphics[height=0.9in]{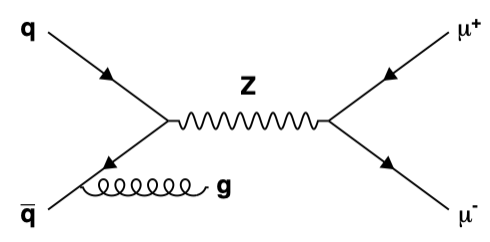}\hfill
\includegraphics[height=0.9in]{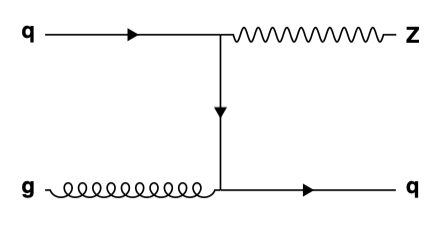}\hfill
\includegraphics[height=0.9in]{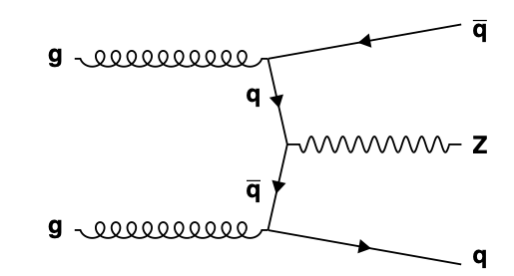}
\caption{Diagrams for \DY production in the \qqbar annihilation process, quark-gluon (\qglu) Compton process, and gluon-gluon (\gluglu) process.}
\label{fig1}
\end{figure}

At Born level, for the quark-antiquark (\qqbar) annihilation process $\qqbar \to \gamma^{*}/Z\to\ell^+\ell^-$ in the \qqbar center of mass, and with the polar axis defined along the incoming quark direction, the angular distribution of the final state negative lepton can be written as
   \begin{eqnarray}
   \label{qqbar}
 \frac{d\sigma}{d\cos\theta} & \propto &  (1 + \cos^{2} \theta) + A_4~\cos\theta.
  \end{eqnarray}
%

For dileptons produced in \pp collisions at the LHC, QCD radiation generally gives the dilepton system nonzero transverse momentum. In this case, the angular distribution can be written more generally in the dilepton rest frame as~\cite{Mirkes1}:
\begin{equation}
\begin{aligned}
\frac{d{\sigma}}{d\cos{\theta}d\phi} \propto &
(1+\cos^2{\theta}) 
+ \frac{1}{2}A_0(1-3\cos^2{\theta})  \\
&+ A_1\sin{2\theta}\cos{\phi} 
+ \frac{1}{2}A_2\sin^2{\theta}\cos{2\phi}  \\
&+ A_3\sin{\theta}\cos{\phi}  
+ A_4\cos{\theta} \\
&+A_5 \sin^2 \theta \sin 2\phi  
+ A_6\sin{2\theta}\sin{\phi} \\
&+ A_7\sin{\theta}\sin{\phi}.
\label{AngleFunc}
\end{aligned}
\end{equation}
The angular coefficients $A_0-A_7$ are functions of the dilepton invariant mass $M_{\ell\ell}$, rapidity $y$, transverse momentum \pt (=\ptZ for dileptons in the Z boson mass region), and parton distribution functions (PDFs). In this paper, the angles $\theta$ and $\phi$ are evaluated in the Collins-Soper (CS) frame, whose definition is summarized in the next section.  

Nonzero \ptZ is balanced by hadronic recoil, so the angular coefficients in Eq.~\ref{AngleFunc} depend on the partonic subprocesses that produce the recoiling parton or jets. The process $\pp\to \DY \to \ell^{+}\ell^-+\text{jets}$ can proceed via three representative subprocesses, as shown in Fig.~\ref{fig1}.
\begin{enumerate} 
\item quark-antiquark annihilation: $\qqbar\to \DY~\Pg$.
\item quark-gluon Compton scattering: $\qglu\to \DY~\Pq$ (or $\Paq\Pg\to \DY~\Paq$).
\item gluon-gluon scattering: $\gluglu\to \DY~\qqbar$.
\end{enumerate}
These subprocesses have different lepton angular distributions, so changes in their relative contributions can change the angular coefficients measured in the inclusive sample.

\begin{figure}[!hbt]
\centering
\includegraphics[width=0.99\columnwidth]{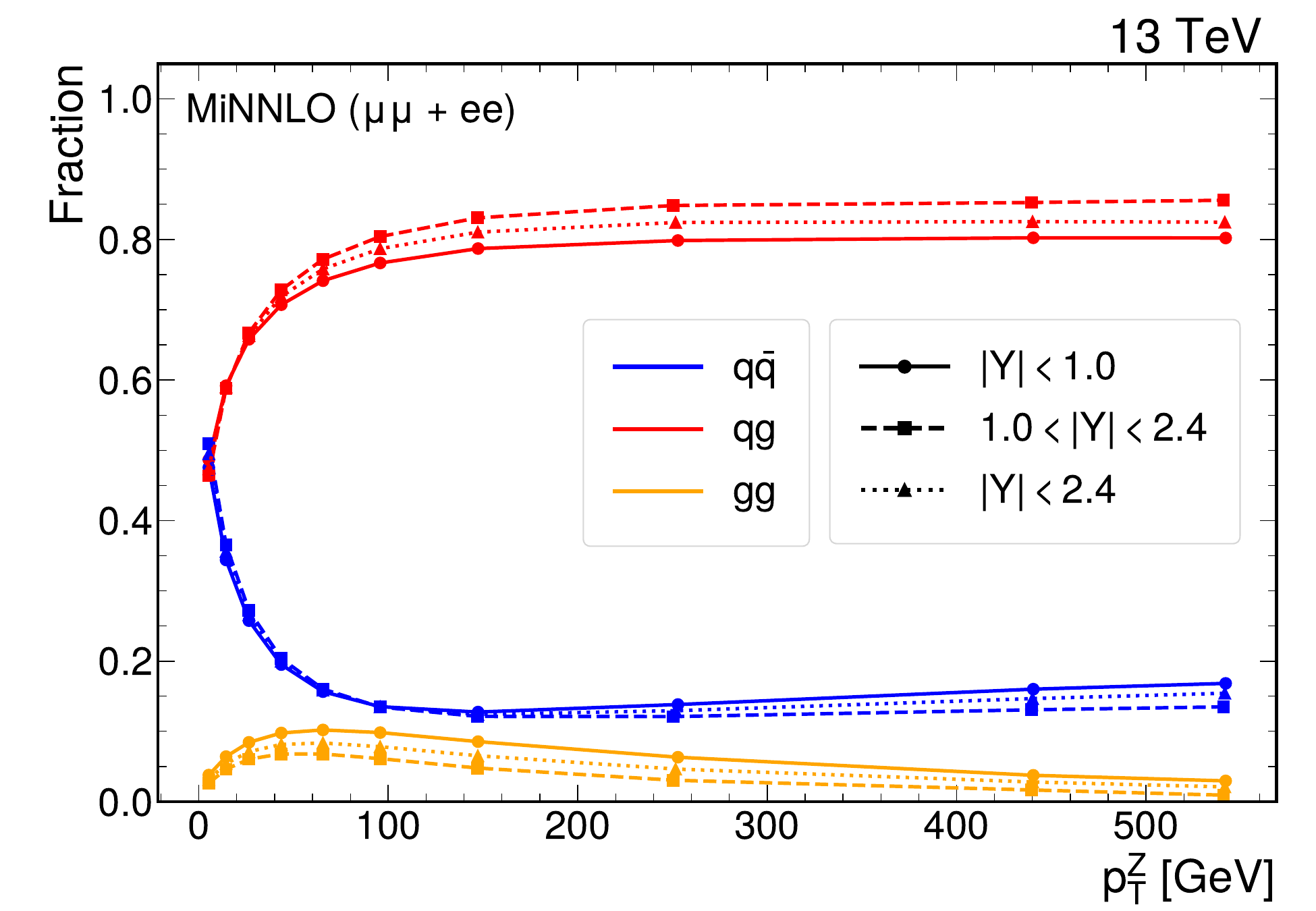}
\caption{Fractional contributions of the \qqbar, \qglu and \gluglu processes in pp collisions at $\sqrt{s}=13\TeV$ as a function of \PZ boson ($81<M_{\ell\ell}<101\GeV$) transverse momentum \ptZ, calculated with the \minnlo MC generator.}
\label{fig2}
\end{figure}

In the \POWHEG-\minnlops event generator~\cite{Monni:2019whf,Monni:2020nks}, referred to as \minnlo below, Fig.~\ref{fig2} shows the fraction of \qqbar, \qglu, and \gluglu events in the \PZ boson region ($80<M_{\ell\ell}<100\GeV$) for \pp collisions at a center-of-mass energy $\sqrt{s} = 13~\TeV$ as a function of \PZ boson transverse momentum \ptZ. The fraction of \qglu events is about half at \ptZ = 0 and increases to 0.8 at high \ptZ, indicating that the relative contributions of the subprocesses vary substantially with \ptZ and make the \ptZ dependence of the inclusive angular coefficients more complex.

In this paper, we use \minnlo predictions for $\pp\to \gamma^{*}/Z \to \ell^+\ell^-$
($ \Pgm^+\Pgm^-$ and $\Pe^{+}\Pe^{-}$)
to study the angular coefficients $A_0$ and $A_2$ for different production subprocesses and different numbers of jets in the final state. We also discuss how the subprocess dependence of these coefficients can be investigated experimentally, including measurements in events with one jet and with one $b$-tagged jet in the final state.

\section{The Collins-Soper reference frame}

\begin{figure}[!hbt]
\begin{center}
\includegraphics[width=3.3in]{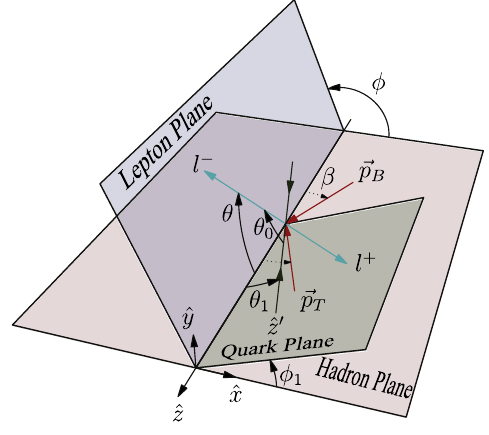}
\caption{Definition of the Collins-Soper frame (from Ref.~\cite{peng3}).
 The hadron plane is formed by $\vec{p}_B$ and $\vec{p}_{T}$, the momentum vectors of the two interacting hadrons. The $\hat{x}$ and $\hat{z}$ axes of the Collins-Soper frame both lie in the hadron plane with $\hat{z}$ axis bisecting the $\vec{p}_B$ and $-\vec{p}_{T}$ vectors. The quark (\Pq) and antiquark (\Paq) annihilate collinearly with equal momenta to form \DY or a vector quarkonium, while the quark momentum vector $\hat{z}'$ and the $\hat{z}$ axis form the quark plane.
The polar and azimuthal angles of $\hat{z}'$ in the Collins-Soper frame are
$\theta_1$ and $\phi_1$. The $\ell^-$ and $\ell^+$ are emitted back-to-back with $\theta$ and $\phi$ as the polar and azimuthal angles for $\ell^-$.}
\label{fig3}
\end{center}
\end{figure}

There are several definitions of the direction of the $z$ axis in the dilepton center of mass frame. 
The most widely used frame is the Collins-Soper (CS)~\cite{Collins:1977iv} frame, shown in Fig.~\ref{fig3}, which is used in this paper.

In the laboratory frame $\theta^{\text{lab}}$ is the polar angle of a particle relative to
the direction of one of the proton beams (the $+z$ axis), $\phi^{\text{lab}}$ is the azimuthal angle, and the pseudorapidity is $\eta=-\ln\left [ \tan(\theta^{\text{lab}}/2)\right ]$.
For the $\ell^{+}\ell^{-}$ pair
 $\pt=p^{\text{lab}}\sin\theta^{\text{lab}}$, $E_{\text{T}}=E^{\text{lab}}\sin\theta^{\text{lab}}$, 
 and the rapidity is 
$y=\frac{1}{2} \ln\frac{E^{\text{lab}}+p_z^{\text{lab}}}{E^{\text{lab}}-p_z^{\text{lab}}}$, 
where $p^{\text{lab}}$ and $p_z^{\text{lab}}$ are the magnitude and $z$ component of the momentum, and $E^{\text{lab}}$ is the energy of the $\ell^{+}\ell^{-}$ pair.

For pp collision in the laboratory frame, the $z$ axis is oriented along the
direction of the incident proton. The transverse component of any vector,
such as the momentum vector, is defined to be relative to the
$z$ axis. For colliding beams (e.g LHC) the beam particle (i.e. positive $z$ axis) is defined as the proton beam that points in the direction of the rapidity of the dilepton pair. 

In the CS frame and all other frames,
the transverse component of vectors
in those reference frames is defined to be relative to
the $z$ axis in those frames.
The polar and azimuthal angles of the $\ell^-$ direction in
the rest frame of the dilepton are denoted as $\theta$ and
$\phi$, respectively.

The ideal positive $z$ axis coincides with the
direction of the incoming quark so that the definition of $\theta$
parallels the definition used in $\Pe^+\Pe^-$ collisions at LEP.
This frame is approximated by the CS rest
frame~\cite{Collins:1977iv} for \pp collisions.

The dilepton rest frame is reached from the laboratory frame via two
Lorentz boosts, first along the laboratory $z$ axis into a frame
where the $z$ component of the lepton-pair momentum vector is zero,
followed by a boost along the transverse component of the lepton-pair
momentum vector. Within the CS frame, the $z$ axis for the polar angle is
the angular bisector between the proton direction and the reverse
of the other proton direction

In the CS frame, we define the momentum vector of the beam particle as $\vec{p}_A$ and the momentum vector of the target particle as $\vec{p}_B$.
The $z$ axis bisects the beam particle direction
and the opposite of the target particle direction in the dilepton rest frame.

The angles $\theta_{CS}$ and $\phi_{CS}$ are defined by 
\begin{equation}
\begin{aligned}
  \cos \theta_{CS} &=\frac{p_z(\ell\ell)}{|p_z(\ell\ell)|} \frac{2}{M_{\ell\ell}\sqrt{{M_{\ell\ell}}^{2} + \ptsq}} (p_{1}^{+}p_{2}^{-} - p_{1}^{-}p_{2}^{+}) \\
    \tan \phi_{CS} &= \frac{\sqrt{M^{2}_{\ell\ell}+\ptsq}}{M_{\ell\ell}} \cdot \frac{\vec{\Delta}\cdot \rthat}{\vec{\Delta}\cdot \pthat}\,.
  \label{CScost}
\end{aligned}
\end{equation}
Here, $p^{\pm}$ corresponds to $\frac{1}{\sqrt{2}}(p^{0}\pm p^{3})$, $p_{1}$ and $p_{2}$ are the 
four-momentum of negatively and positively charged leptons, respectively, 
\pt is the transverse momentum of the dimuon pair in the laboratory system. The three components of the vector $\vec{\Delta}$ are 
$\Delta^{j}=p_1^{j}-p_2^{j}$, ~\pthat is a transverse unit vector in the direction of $\vecpt$, and \rthat is a transverse unit vector in the direction of $\vec{p}_A\times \vecpt$.
  
The angle $\phi_{CS}$ is the angle between the direction of the
\DY boson \pt and the direction of the negatively charged lepton.
%
\section{Expectation for angular coefficients in the Collins-Soper frame}
For the $q\bar{q}\to \gamma^{*}/Z~g$ annihilation process~\cite{qqCollins:1978yt,bnlBoer:2006eq,berger1,berger2,bodek}, 
perturbative QCD at leading order (LO) predicts that the angular coefficients 
$A_0$ and $A_2$ are equal,
independent of Parton Distribution Functions (PDFs) or of the rapidity $y$,
and are only functions of the ratio $\pt/{M_{Z}}$ 
as given by 
 \begin{equation}
 A_0^{\qqbar} =   A_2^{\qqbar} = \frac{\ptsq}{{M_{Z}}^2+\ptsq} \,.
\label{predictQ}
\end{equation}
For the \qqbar process, the above expression for $A_0^{\qqbar}$ can be derived~\cite{bodek} from simple geometrical arguments at leading order. The geometric derivation holds whenever the vector-boson transverse momentum is generated entirely by radiation from one initial-state parton, with the other parton non-radiating, in this single-emission configuration the relation is preserved to all orders. In addition, Refs.~\cite{berger1,berger2} show that the same relation is preserved to all orders for the leading logarithmically enhanced, resummed contribution.

For the $\qglu \to \DY~\Pq$ Compton process, $A_0$ and $A_2$ 
depend on PDFs and $y$. Based on the work 
of Refs.~\cite{qqCollins:1978yt,gq2Lindfors:1979rc,gq3Thews:1979rr,gq4Noman:1978eh,NA10:1986fgk}
in perturbative QCD at LO, when averaged 
over $y$, $A_0^{\qglu}$ and $A_2^{\qglu}$ are approximately~\cite{NA10:1986fgk}
described by 
\begin{eqnarray}
A_0^{\qglu} &=& A_2^{\qglu} \approx \frac{5\ptsq}{{M_{Z}}^2+5\ptsq}
\label{predictG}
\end{eqnarray}
%
\subsection{The Lam-Tung relation}
The equality $A_2= A_0$ is known as the Lam-Tung (L-T) relation~\cite{LTLam:1978zr}. At LO, it is valid for 
both \qqbar and \qglu processes~\cite{bnlBoer:2006eq}. At higher orders $A_2$ is smaller than $A_0$ because the emission of additional partons distorts the $\phi$ distribution (smearing). Therefore, if 
events with two jets or more are not included in the sample, the violation of the L-T relation is expected to be smaller.
%
%
\begin{figure*}[hbt]
\begin{center}
\includegraphics[width=0.95\textwidth]{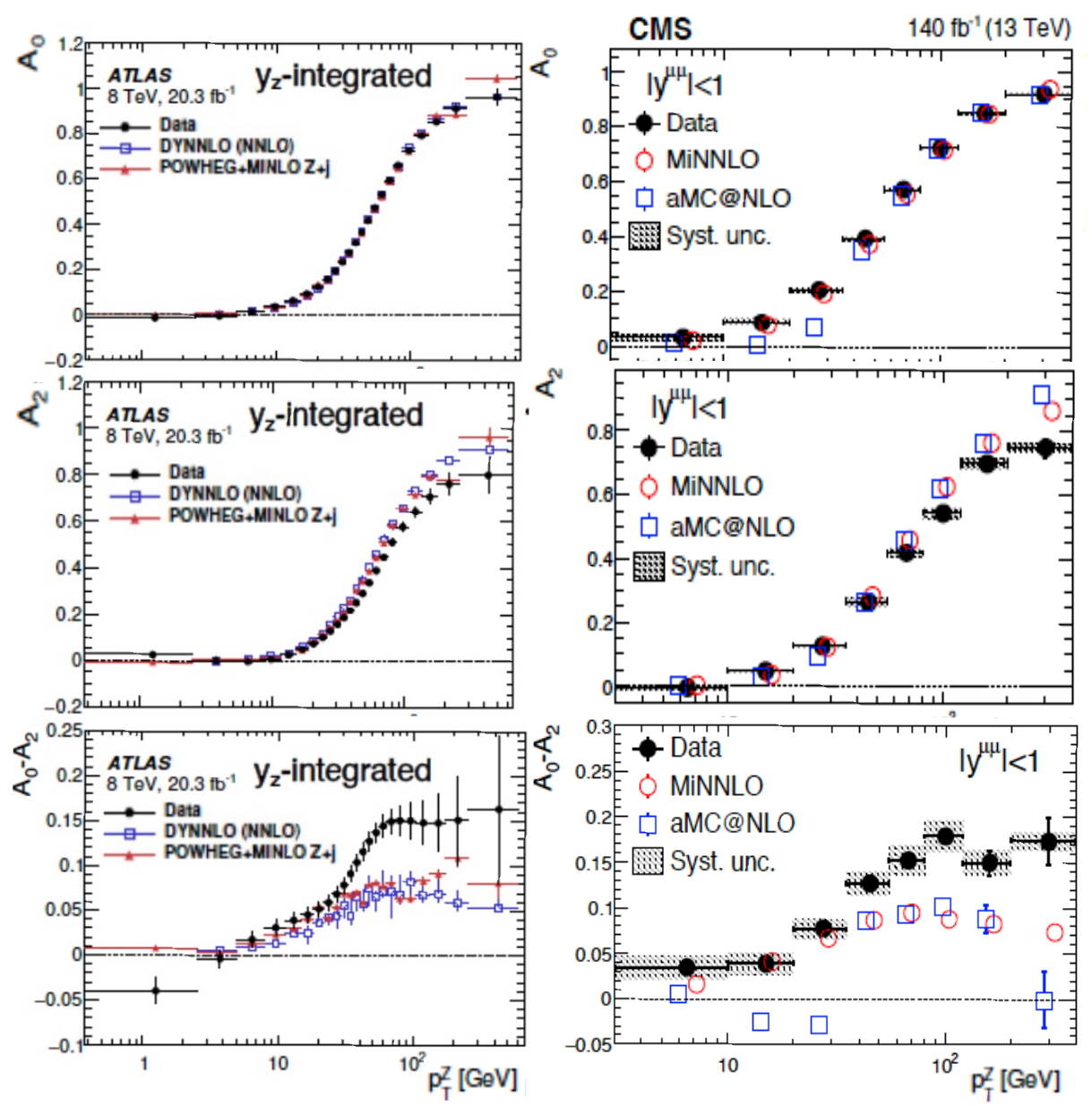}
\caption{Some of the ATLAS measurements~\cite{ATLASang} at 8\TeV and CMS measurements~\cite{CMS:2026amb} at 13\TeV. The top two panels  show that both  measurements of $A_0$ are in good agreement with the predictions of \POWHEG + \minlo~\cite{Hamilton:2012np}. The two  middle panels  show that both  measurements of $A_2$ are lower than the predictions of \POWHEG + \minlo above \ptZ of 20\GeV. And the bottom two panels show that $A_0-A_2$ is non-zero in violation of the L-T relation in both the ATLAS 8\TeV and CMS 13\TeV measurements. The measured violations are larger than theory predictions. Figures from \cite{ATLASang} and \cite{CMS:2026amb}. }
\label{fig4}
\end{center}
\end{figure*}
%
\section{Measurements of \PZ boson angular coefficients}
The first measurement (published in 2011~\cite{CDF:2011ksg}) of \PZ boson angular coefficients at a hadron collider was performed by the CDF collaboration in proton-antiproton collisions at $\sqrt{s}=1.96\TeV$, using $2.1\,\mathrm{fb}^{-1}$ of data.  Within its experimental precision, the CDF measurement found agreement with perturbative QCD predictions and with the Lam--Tung (L-T) relation. More precise measurements in \pp collisions at 8\TeV were published by CMS in 2015~\cite{CMSang}, by ATLAS at 8\TeV in 2016~\cite{ATLASang}, by CMS at 13\TeV in 2026~\cite{CMS:2026amb} and  by LHCb at 13 TeV in 2022\cite{LHCb:2022tbc}. The more precise LHC measurements observe violation of the L-T relation in the \PZ boson mass region as discussed below.
\begin{figure}[tb]
\includegraphics[width=\columnwidth]{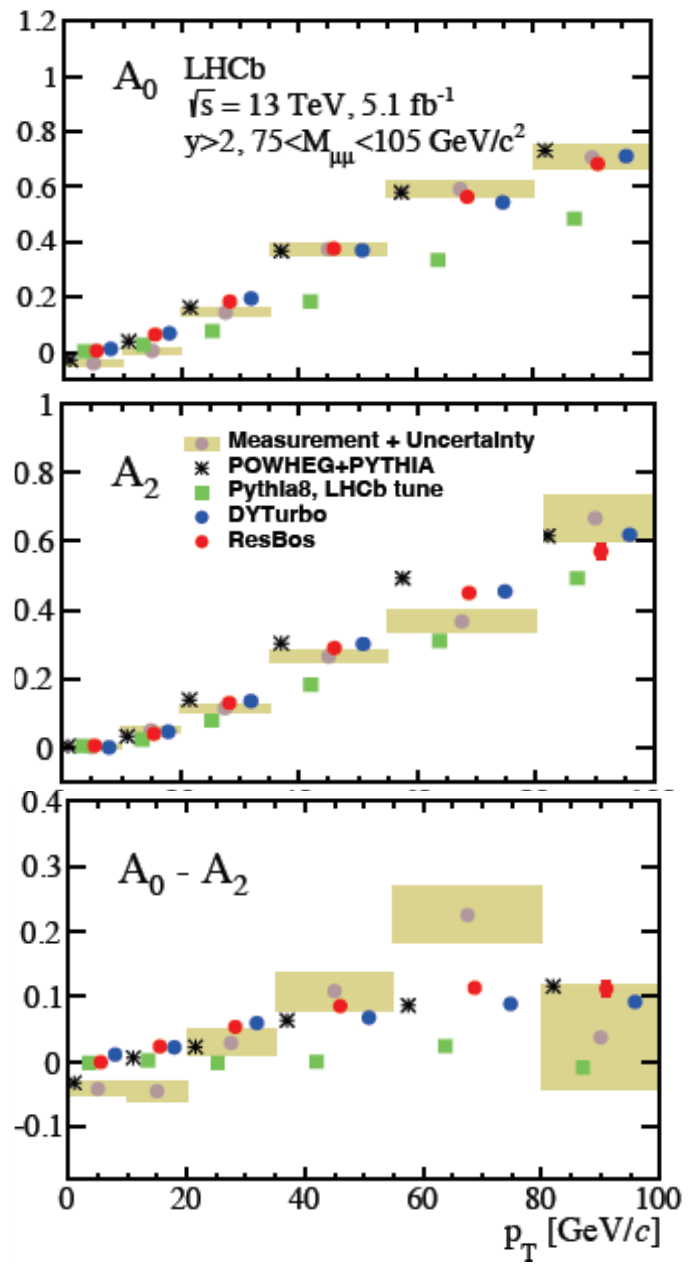}
\caption{Comparison of LHCb measurements of $A_0$ and $A_2$ with different predictions, as a function of the \ptZ for Z-boson rapidity $y^Z>2$. The total uncertainty (shown in the figure) is dominated by the statistical component. The theoretical predictions correspond to the same Z-boson \ptZ bins as data and the \ptZ shifts of the theoretical markers in all plots are for visualization purposes only. {\textsc{DYTurbo}}\cite{dyturbo} and {\textsc{resbos}}\cite{resbos} predictions include the theoretical uncertainties.  Figure from \cite{LHCb:2022tbc}. }
\label{Fig5}
\end{figure}

Figure~\ref{fig4} shows some of the measurements of $A_0(\ptZ,y)$, $A_2(\ptZ,y)$ and $A_0-A_2$ by  ATLAS~\cite{ATLASang}  at 8\TeV (left three panels)  and CMS~\cite{CMS:2026amb} at 13\TeV (right three panels). 
The top two panels  show that both  measurements of $A_0$ are in good agreement with  \POWHEG + \minlo~\cite{Hamilton:2012np} predictions.

In contrast, the  middle two panels show that both  measurements of $A_2(\ptZ,y)$ are lower than the predictions above \ptZ of 20\GeV. The bottom two panels show that $A_0-A_2$ is non-zero in violation of the L-T relation in both the ATLAS 8\TeV and CMS 13\TeV measurements. In addition,  the measured violation is larger than all theory predictions.  Similar plots for LHCb\cite{LHCb:2022tbc} are shown in Fig.\ref{Fig5}. 

Existing measurements have not separated the contributions of different partonic subprocesses or investigated the jet-multiplicity dependence of the L-T violation. We therefore investigate these dependencies using MC simulations in the sections below.
\section{Monte Carlo studies at 13\TeV}
In this paper we use Monte Carlo (MC) generated events combining the $\mu\mu$ and $ee$ channels to investigate these issues, and propose how they can be investigated experimentally.

We use the \POWHEG v2 event generator with \minnlops v4.0 (at next-to-next-to-leading order, NNLO, accuracy in QCD)~\cite{Monni:2019whf, Monni:2020nks}, matched to \PYTHIA 8.2 for parton showering~\cite{Sjostrand:2014zea}. The NNPDF3.1 NNLO parton distribution function set is used~\cite{NNPDF:2017mvq}. Photon final-state radiation is simulated with the \PHOTOS 3.61 package~\cite{Barberio:1993qi, Golonka:2005pn}. 
The $\pp\to\DY\to\ell^+\ell^-$ MC sample is simulated with a center-of-mass energy of 13\TeV (corresponding to Run 2 at the LHC).. 
\vspace{-0.1in}
\section{Extraction of angular coefficients from MC event samples}
%

\begin{figure*}
\begin{center}
\includegraphics[width=3.0in,height=2.2in]{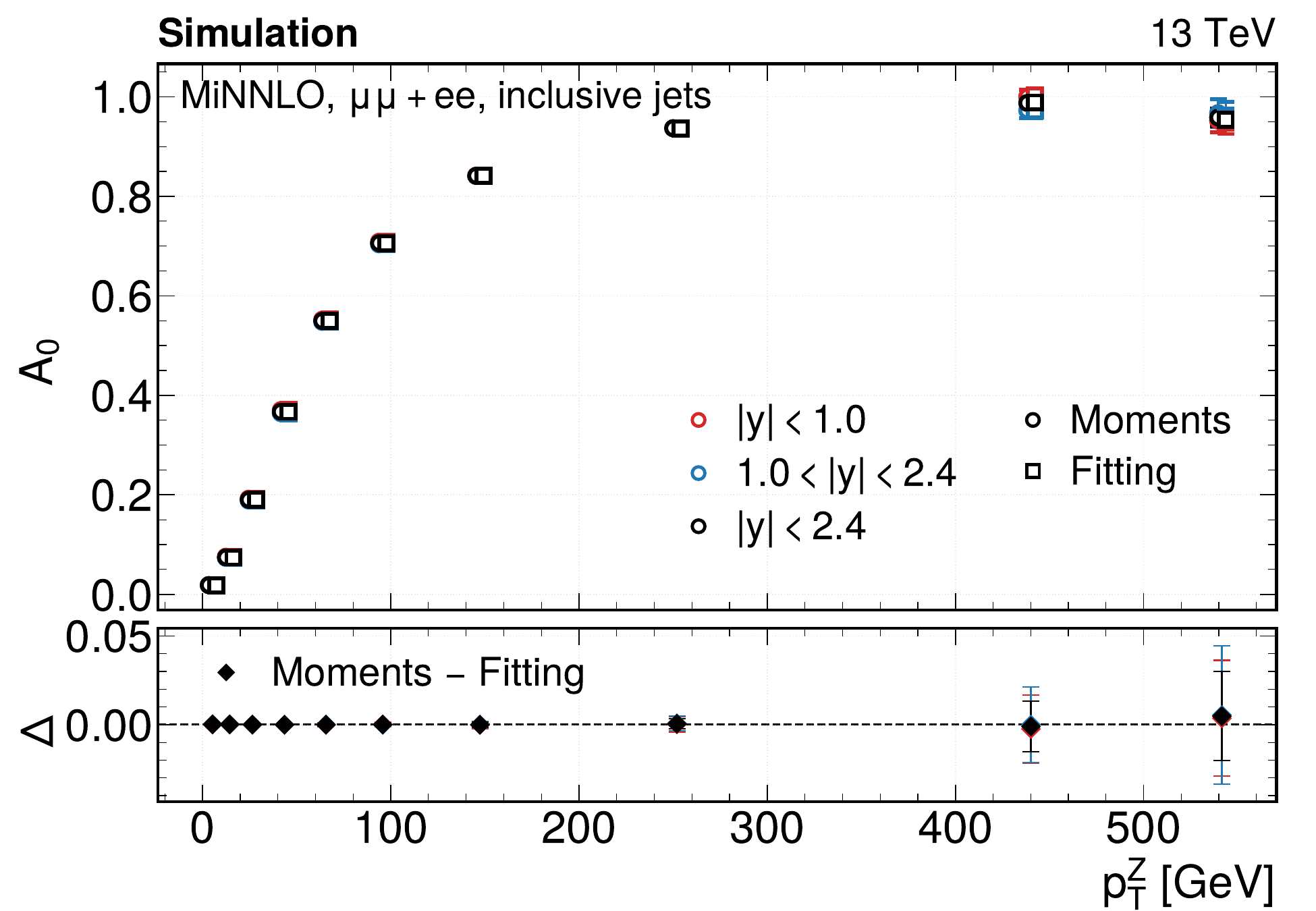}
\includegraphics[width=3.0in,height=2.2in]{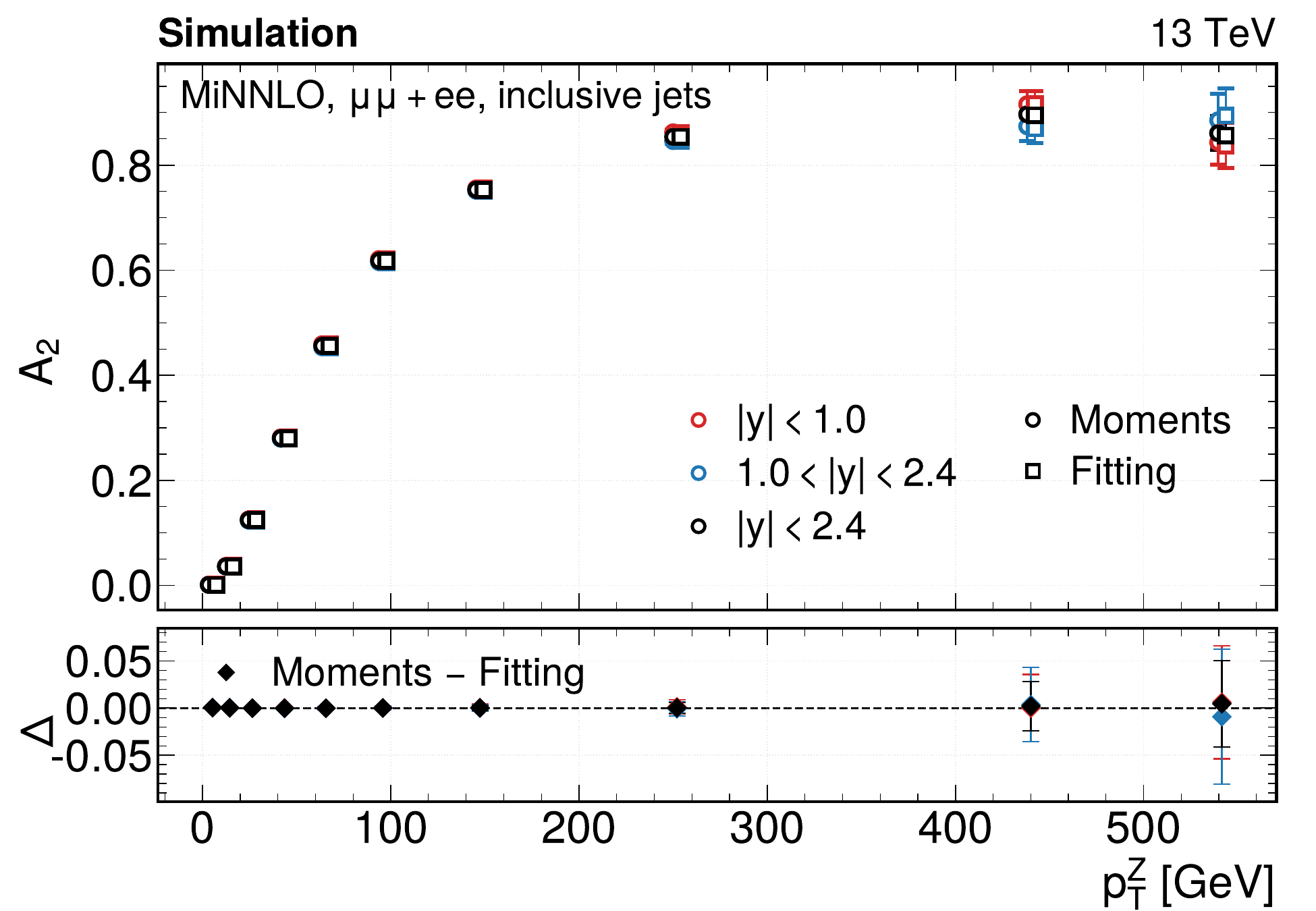}
\includegraphics[width=3.0in,height=2.2in]{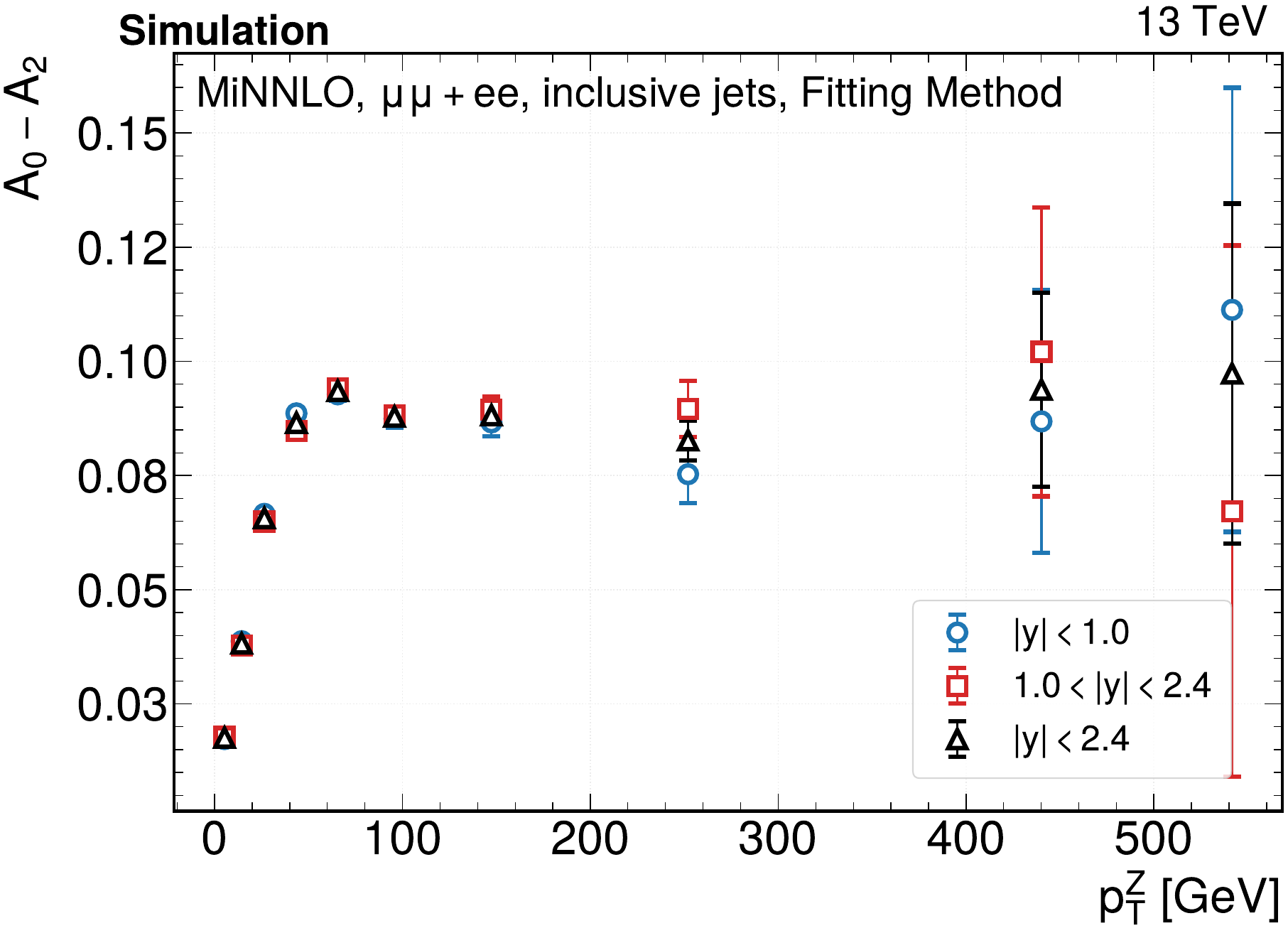}
\includegraphics[width=3.0in,height=2.2in]{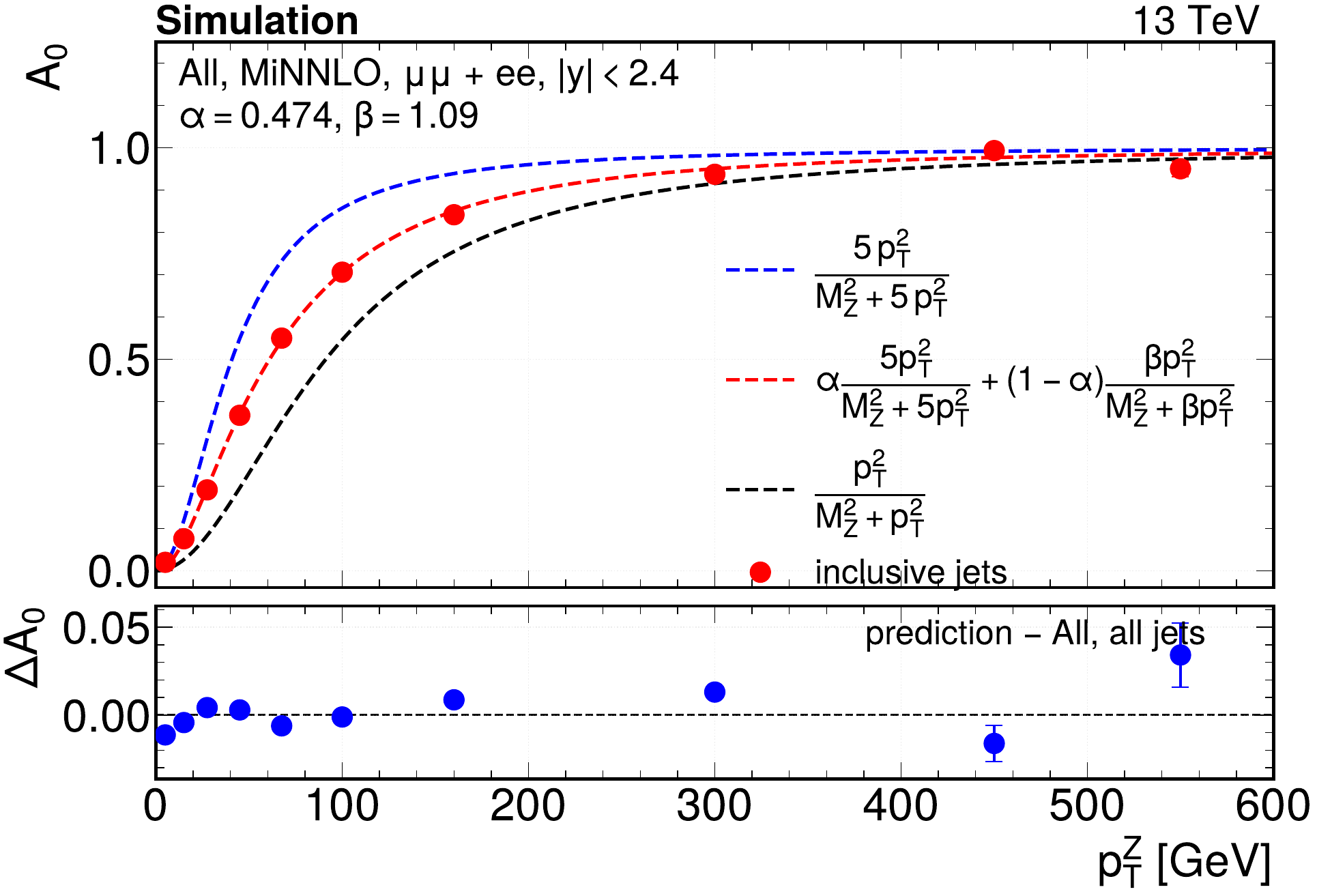}
\caption{Top two panels: Comparison of moment and fit method for the measurements of $A_0 (\ptZ,y)$ (left) and $A_2 (\ptZ,y)$ (right)
for events with $|\yZ|<1$, $1<|\yZ|<2.4$ and $|\yZ|<2.4$. 
Bottom left panel: ($A_2 (\ptZ,y)-A_0(\ptZ,y)$) for events with $|\yZ|<1$, $1<|\yZ|<2.4$ and $|\yZ|<2.4$. Bottom right panel: Comparison of fit and MC for the $\ptZ$ dependence of $A_0 (\ptZ,y)$ for events with $0<|\yZ|<2.4$.
}
 \label{Fig6}
\end{center}
\end{figure*}
\vspace{-0.1in}
The theory predictions for the angular coefficients are extracted by fitting the 
MC generated angular distributions directly.

When integrated over $\phi$ the differential cross section reduces to:
\begin{eqnarray}
\frac{d{\sigma}}{d\cos{\theta}} \propto (1+\cos^2{\theta})
+ \frac{1}{2}A_0 (1-3\cos^2{\theta}) + A_4 \cos{\theta}
\label{etheta4}
\end{eqnarray}
Here, the angular distribution is then only a function of $A_0 (M_{Z}, \pt,y)$ and $A_4 (M_{Z},\pt,y)$.

If we also sum the positive and negative values of $\cos{\theta}$, $A_4$ cancels and the angular distribution is then only
a function of $A_0 (M_{Z}, \pt,y)$.
\begin{eqnarray}
\frac{d{\sigma}}{d|\cos{\theta}|} \propto [1+\cos^2{\theta}]
+ \frac{1}{2}A_0 [1-3\cos^2{\theta}]  
\label{etheta}
\end{eqnarray}
In order to measure $A_2(M_{Z}, \pt,y)$ we sum over positive and negative values of $\cos\theta$, $\cos\phi$, and $\sin\phi$, which cancels the terms odd in $\cos\theta$ or $\phi$ and leaves only the $A_0$ and $A_2$ terms. We then obtain
\begin{eqnarray}
\frac{d{\sigma}}{d|\cos{\theta}|d\phi}& \propto& [1+\cos^2{\theta}]
+ \frac{1}{2}A_0 [1-3\cos^2{\theta}]\nonumber\\
&&+\frac{1}{2}A_2\sin^2\theta (1-2\sin^2\phi)
\label{ethetaphi}
\end{eqnarray}
The two-dimensional angular distribution in $\cos^2{\theta}$ and $\sin^2\phi$ is a function of both $A_0(M, \pt,y)$ and $A_2(M, \pt,y)$.
Here both $A_0(M, \pt,y)$ and $A_2(M, \pt,y)$ are extracted from fits to the two dimensional distribution of $\cos^2{\theta}$ and $\sin^2\phi$. 
%
\begin{figure}[t]
\begin{center}
\includegraphics[width=\panelwidth]{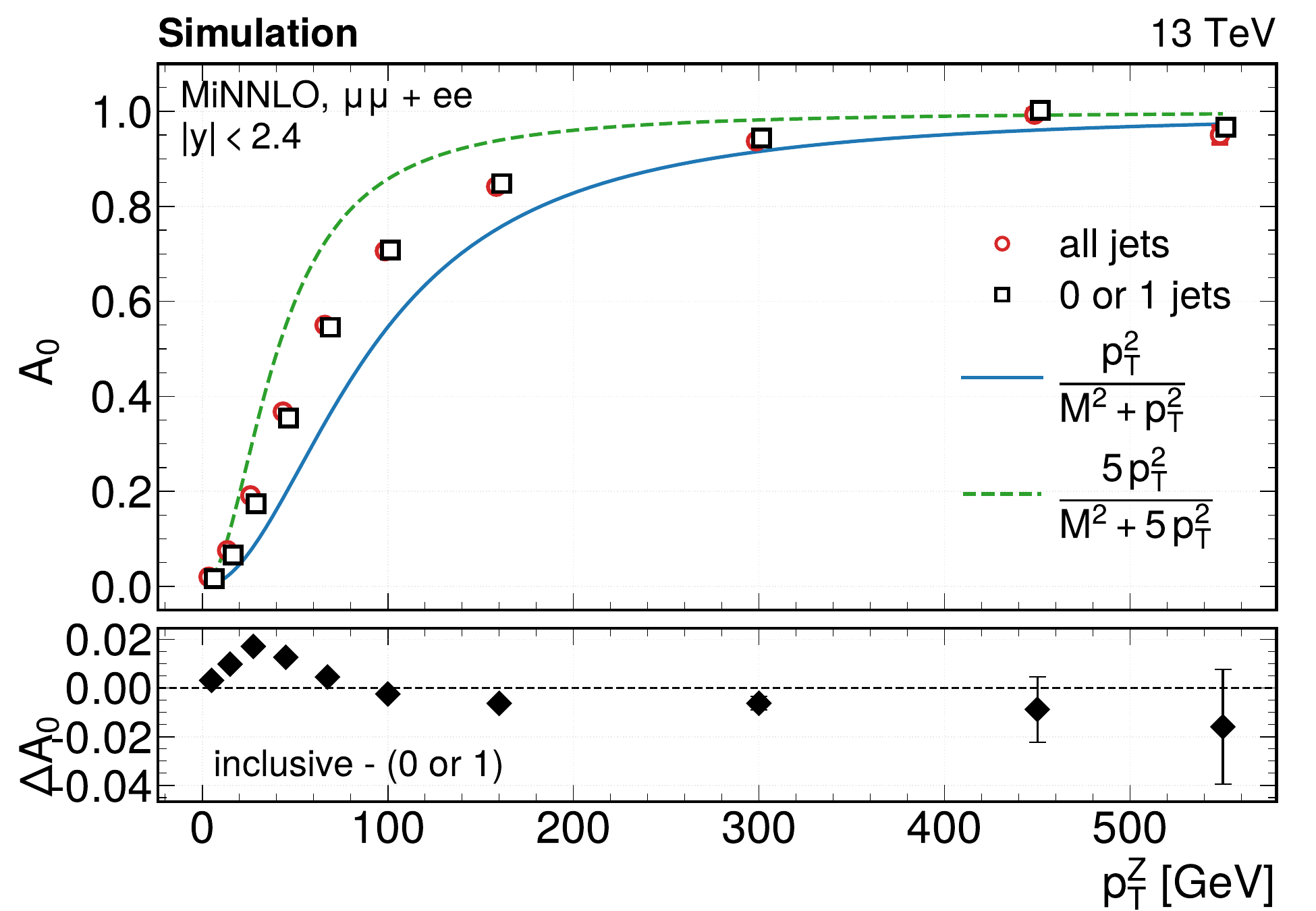}
\includegraphics[width=\panelwidth]{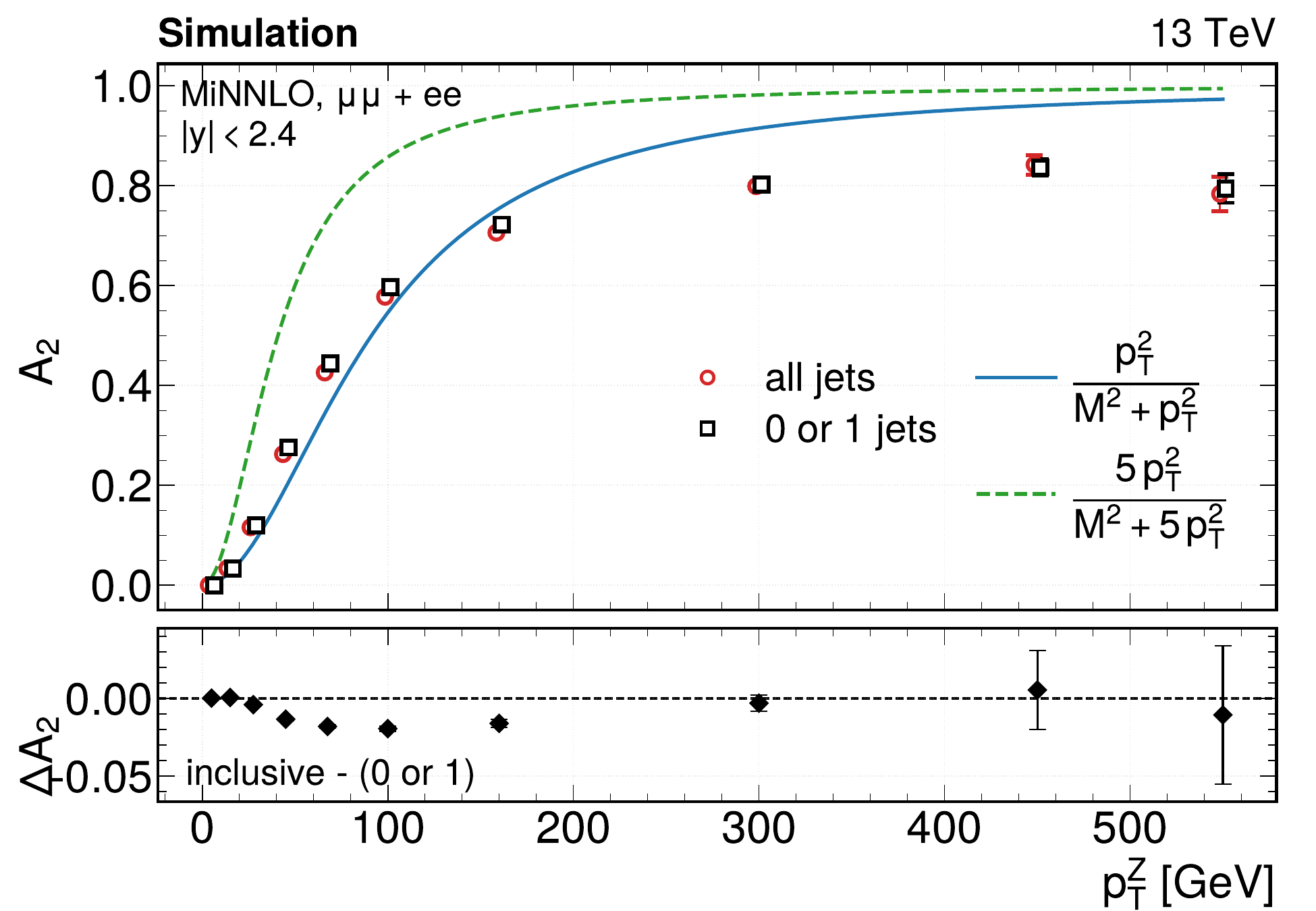}
\includegraphics[width=\panelwidth]{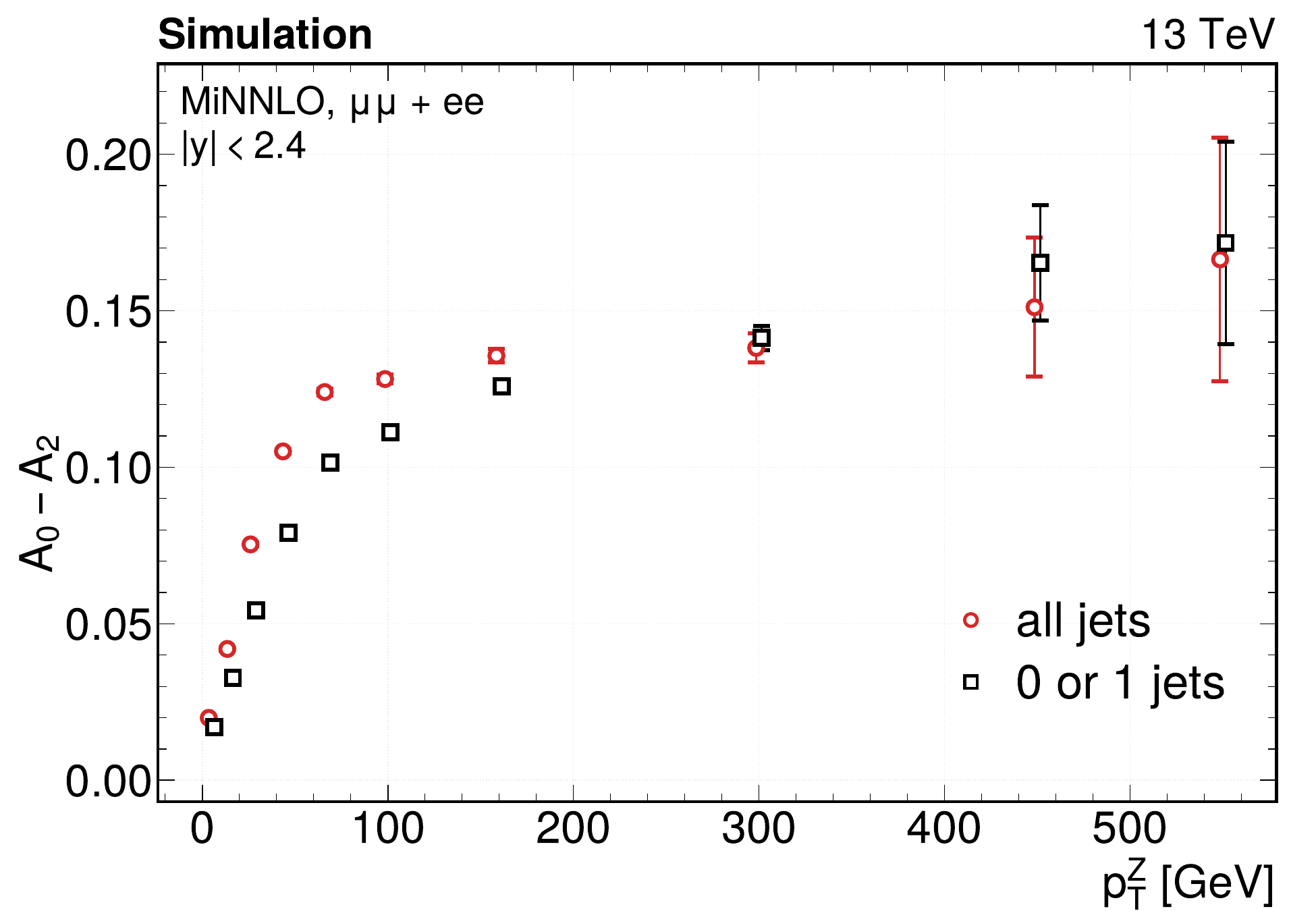}
\caption{ Comparison of $A_0(\ptZ)$ and $A_2(\ptZ)$, and $[A_0(\ptZ)-A_2(\ptZ)]$
for all events versus events with $\le$ 1 jet (with $\ptjet > 20\GeV$, $|\yZ|<2.4$ and $|\eta^{\text{jet}}|<2.4$).}
 \label{Fig7}
\end{center}
\end{figure}

Averaging over the \PZ boson mass range ($81<{M_{\ell\ell}}<101\GeV$) we use Eq.~\ref{etheta} 
to extract $A_0(\ptZ)$ and Eq.~\ref{ethetaphi} to extract $A_2(\ptZ)$ for data in a specific range of $|y|$. 

Note that if we integrate over $\cos^2{\theta}$ the dependence on $\cos^2{\theta}$ integrates to 1.0 and we obtain $\frac{d{\sigma}}{d\phi} \propto 1 + \frac{1}{4}A_2 (1-2\sin^2\phi)$. 
In principle, this expression can be used to extract $A_2(\ptZ)$. However, since we are dealing with a finite statistical sample, the integration over $\cos^2{\theta}$ 
may not integrate to exactly 1.0. Consequently, we perform a two-dimensional fit using Eq.~\ref{ethetaphi} to extract $A_2(\ptZ)$.

An alternative method of extracting the angular coefficients is the method of moments described in Appendix~\ref{appendix:moments}. The direct fitting of the angular distributions method and the method of moments can only be used on generator level samples with no fiducial acceptance cuts. 
In order to extract $A_0$ and $A_2$ in reconstructed MC and experimental data in a collider experiment template fitting is used. In template fitting the reconstructed angular distributions in $\cos\theta$ and $\phi$ are compared to MC predictions in which the generator level distributions are re-weighted to obtain reconstructed distributions in $\cos\theta$ and $\phi$ for different generator level values of $A_0(\ptZ)$ and $A_2(\ptZ)$ in each bin of \ptZ, and the extracted values of $A_0$ and $A_2$ are the values  that minimize the $\chi^2$.
\vspace{-0.25in}
\section{MC Studies of $A_0(\ptZ)$ and $A_2(\ptZ)$} 
\subsection{Rapidity dependence of $A_0$ and $A_2$}
As shown in the first three panels of Fig.~\ref{Fig6} the values of $A_0(\ptZ)$, $A_2(\ptZ)$ and $[A_0(\ptZ)-A_2(\ptZ)]$  are consistent within their uncertainties for $|\yZ|<1$, $1<|\yZ|<2.4$,  and the inclusive rapidity range $|\yZ|<2.4$. Therefore, in the remainder of this paper results are shown only for the inclusive rapidity sample, to maximize statistical power. Additionally, the first three panels of  Fig.~\ref{Fig6}, confirm that the values of  $A_0(\ptZ)$ and $A_2(\ptZ)$ extracted via direct fitting and the method of moments are consistent.

\subsection{Fitting the $\ptZ$ dependence of $A_0(\ptZ)$}
The angular-weighting method~\cite{bodek} for the precise extraction of the electroweak mixing angle from the angular coefficient $A_4$ requires a good parameterization of $A_0(\ptZ)$. 
The values of $A_0(\ptZ)$ for $0<|\yZ|<2.4$ (shown on the bottom right panel of Fig.~\ref{Fig6}) are well described by the following functional form:
\begin{equation}
\label{A0_Fit}
A_0(\ptZ)=
\alpha
\left[
\frac{5\ptsq}{{M_{\ell\ell}}^2+5\ptsq}
\right]
+
(1-\alpha)
\left[
\frac{\beta\,\ptsq}{{M_{\ell\ell}}^2+\beta\,\ptsq}
\right],
\end{equation}
with $\alpha=0.474$ and $\beta=1.09$ 
(shown as the dashed red line in the bottom right panel of Fig.~\ref{Fig6}).

Similarly, $A_0(\ptZ)$ for $0<|\yZ|<1.0$ is described by the same form
with $\alpha=0.485$ and $\beta=1.08$, and $A_0(\ptZ)$ for $1.0<|\yZ|<2.4$
is described by the same form with $\alpha=0.466$ and $\beta=1.1$.
\vspace{-0.2in}
\subsection{Violation of the Lam-Tung relation}
The ATLAS, CMS and LHCb  measurements at the LHC show that for an inclusive sample of events 
 $A_2(\ptZ)$ is smaller than $A_0(\ptZ)$ thus violating the
L-T relation. Our MC studies described below show that this violation originates from events with more than one hadronic jet in the final state (since the second jet smears the angular distribution in $\phi$).

Experimentally, final state hadronic jets can be reliably measured~\cite{CMS:2026mec} for $\ptjet > 20\GeV$ (b-quark tagging in jets can be measured more reliably for $\ptjet > 30\GeV$).   Therefore we only investigate final state hadronic jets with $\ptjet > 20\GeV$. 
%
%
The top panel of Fig.~\ref{Fig7} shows $A_0(\ptZ)$ for the inclusive Z sample compared to $A_0(\ptZ)$ for a sample of \PZ boson events with $\le$ 1 jet in the final state (with $\ptjet > 20\GeV$ and $|\eta^{\text{jet}}|< 2.4$). The values of $A_0(\ptZ$) for the inclusive sample and the sample of events with $\le$ 1 jet are close to each other.
  
The middle panel of Fig.~\ref{Fig7} shows the same comparison
for $A_2(\ptZ)$ for the two categories of events. Here, the values of $A_2(\ptZ)$ for events with $\le$ 1 jet in the final state are higher than the values for the inclusive sample (which includes events with two or more jets with $\ptjet > 20\GeV$).

The bottom panel of Fig.~\ref{Fig7} shows the same comparison for the difference [$A_0(\ptZ)-A_2(\ptZ)$] for the two categories.
The violation of the L-T relation in the MC sample is smaller for events with $\le 1$ jet 
than for the inclusive sample of events. This can be verified experimentally with the  Drell-Yan data samples at the LHC.

In conclusion, our MC study indicates that the violation of the L-T relation originates from multijet events. The fact that in the ATLAS 8\TeV and CMS 13\TeV data the violation of the L-T relation is larger than in the MC indicates that multijet events are mismodeled in the MC. 
%
%
%
\vspace{-0.2in}
%
%
\begin{figure}[htbp]
\begin{center}
\includegraphics[width=\panelwidth]{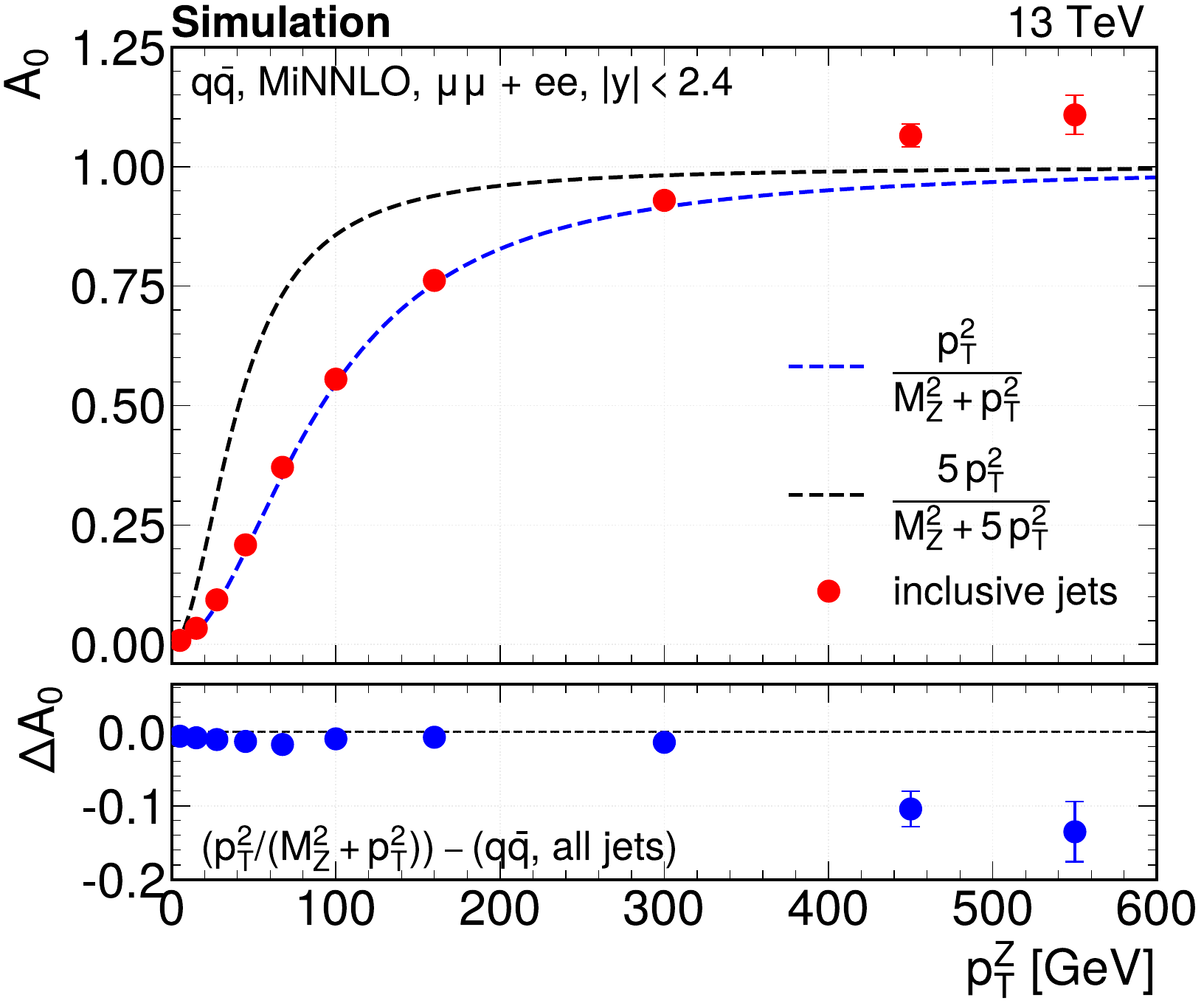}
\includegraphics[width=\panelwidth]{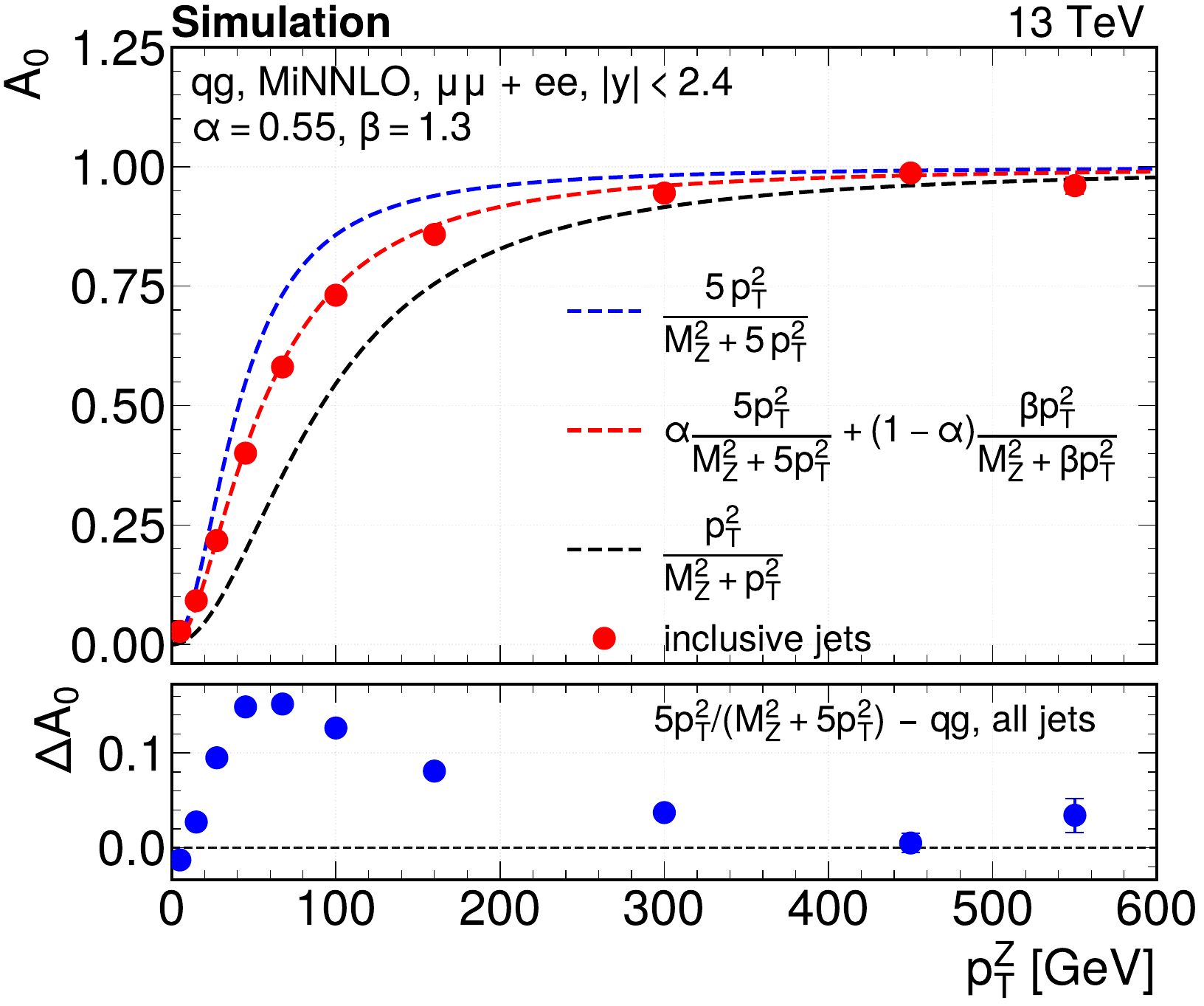}
\includegraphics[width=\panelwidth]{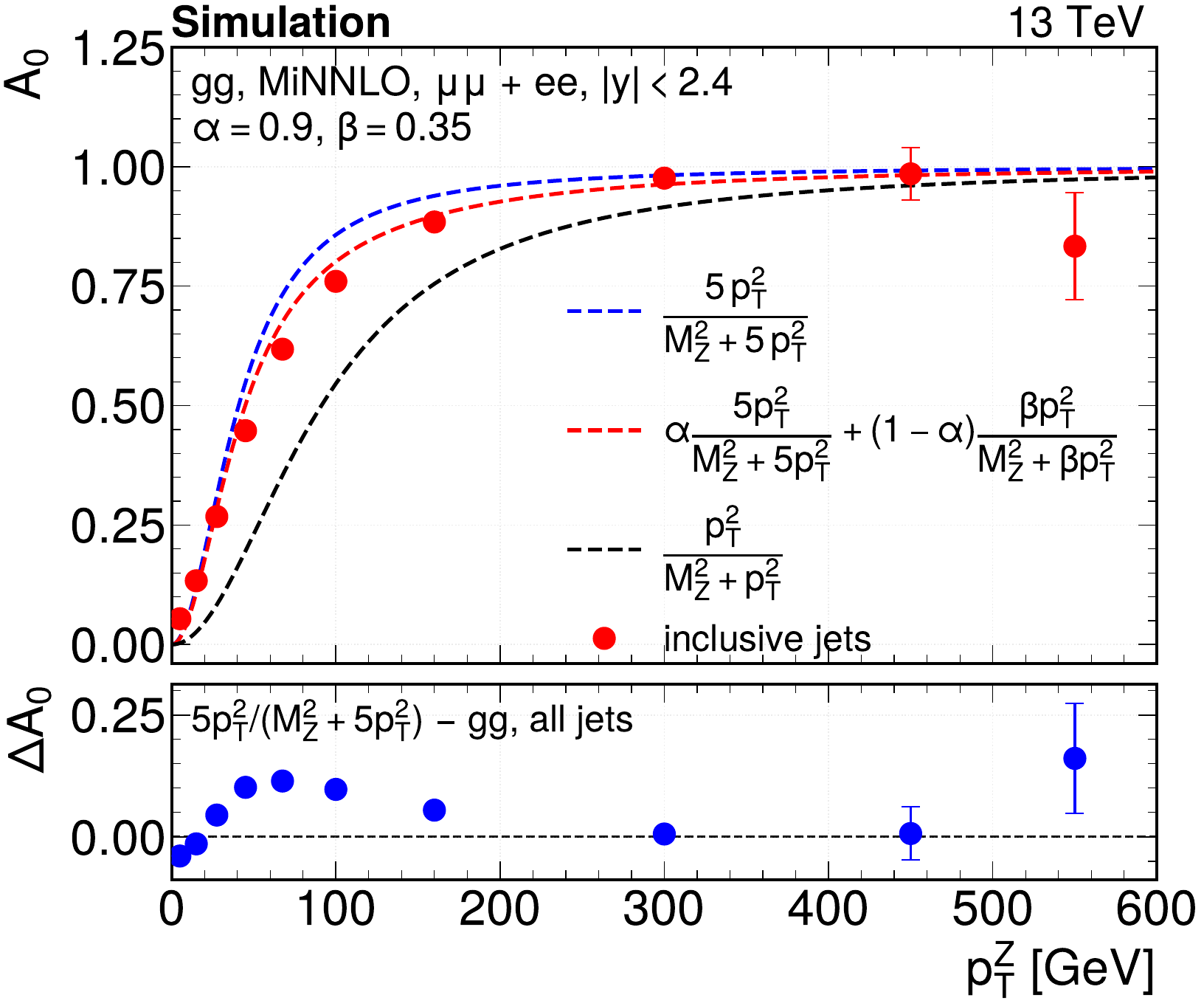}
\caption{ Top panel: The $\ptZ$ dependence of $A_0^{\qqbar}$ compared to $\ptsq/({M_{\ell\ell}}^2+\ptsq)$. Middle panel:The $\ptZ$ dependence of $A_0^{\qglu}$ 
compared to $5\ptsq/({M_{\ell\ell}}^2+5\ptsq)$. Bottom panel: The $\ptZ$ dependence $A_0^{\gluglu}$ compared to
$5\ptsq/({M_{\ell\ell}}^2+5\ptsq)$. In the middle and bottom panels the red dashed lines are our better fits for the $\ptZ$ dependence as described in the text.
}
 \label{Fig8}
\end{center}
\end{figure}
\begin{figure}[htbp]
\begin{center}
\includegraphics[width=\panelwidth]{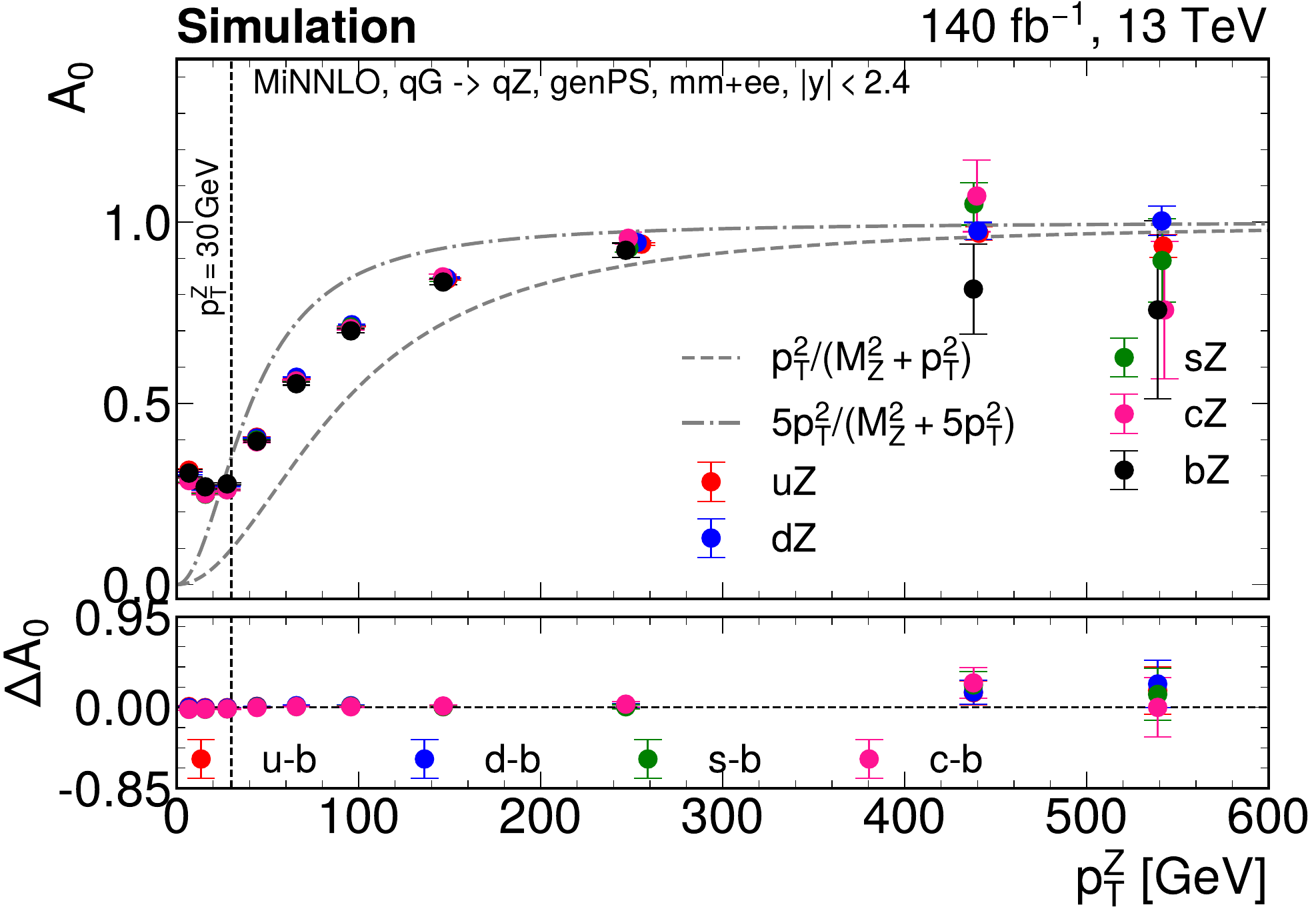}
\includegraphics[width=\panelwidth]{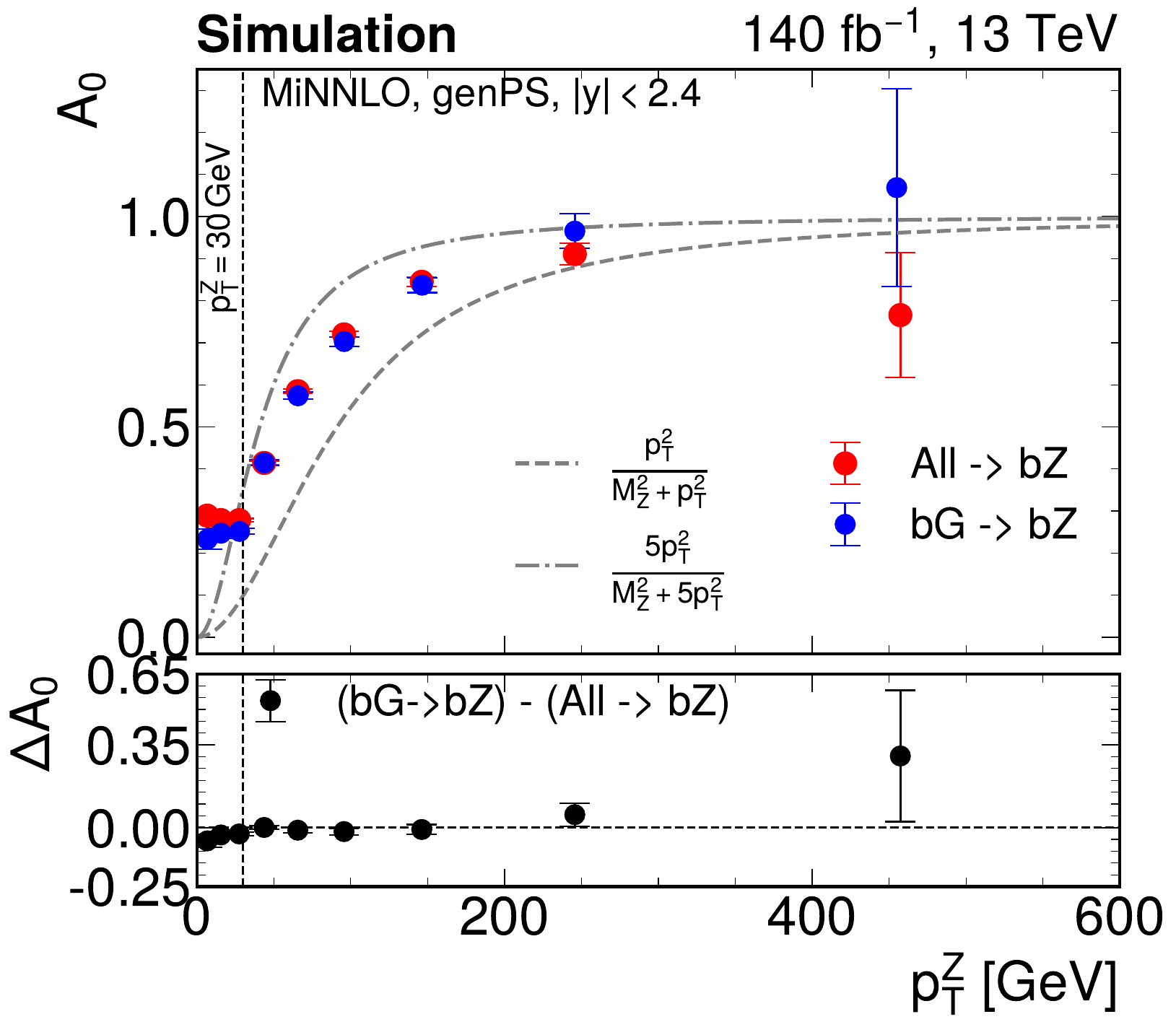}
\caption{Top panel: Comparison of $A_0$ for quark-gluon processes $\qglu \to \Pq\PZ$ for different quark flavors $\Pq=\Pqu,\Pqd,\Pqs,\Pqc$ and $\Pqb$ with one jet in the final state. Bottom panel: Comparison of $A_0$ for $\Pqb\Pg\to \Pqb\PZ$ quark-gluon process to 
\textit{all processes} $\to \Pqb\PZ$ (with one jet in the final state). This indicates that the validity of the \qglu MC prediction can be investigated experimentally in a sample of events with a single $\Pqb$-tagged jet in the final state as such events are produced through the \qglu process.
}
 \label{Fig9}
\end{center}
\vspace{-0.25in}
\end{figure}
\subsection{Comparison of $A_0^{\qqbar}$, $A_0^{\qglu}$ and $A_0^{\gluglu}$}
For the \qqbar process, the angular coefficient $A_0$ is expected from geometrical considerations~\cite{bodek} to follow
\begin{equation}
A_0^{\qqbar} = \frac{\ptsq}{M^2+\ptsq},
\end{equation}
where $M$ is the dilepton invariant mass.
This expression is expected to remain valid to all orders in QCD for configurations in which the \PZ boson recoils against one or more jets emitted by only one initial state quark or antiquark.
Our {\POWHEG}+{\minnlops} simulation, shown in the top panel of Fig.~\ref{Fig8}, indicates that $A_0^{\qqbar}$ is well described by the geometric form for $\ptZ$ less than 200 GeV. 
For the \qglu Compton process, $A_0^{\qglu}$ is larger than $A_0^{\qqbar}$.
Previous theoretical studies~\cite{peng2} approximated $A_0^{\qglu}(\ptZ)$ by the following
expression:
\begin{equation}
A_0^{\qglu}(\ptZ)\approx \frac{5\ptsq}{5\ptsq+{M_{\ell\ell}}^2}.
\end{equation}
Our {\POWHEG}+{\minnlops} simulation, shown in the middle panel of Fig.~\ref{Fig8}, indicates that this approximation overestimates the predicted values of $A_0^{\qglu}$, particularly at high \ptZ. We find that the simulated \qglu contribution is better described by the parametrization given in Eq.~\ref{A0_Fit} (shown as the dashed red line in the middle panel of Fig.~\ref{Fig8}) with $\alpha=0.55$ and $\beta=1.30$.

For the ${\gluglu}$ process, the simulated $A_0^{\gluglu}(\ptZ)$ is also better described by the parametrization given in Eq.~\ref{A0_Fit} with $\alpha=0.9$ and $\beta=0.35$. This fit is shown as the red dashed line in the bottom panel of Fig.~\ref{Fig8}.
\section{Using events with a single $\Pqb$ tagged jet to measure $A_0^{\qglu}$}
The top panel of Fig.~\ref{Fig9} shows a comparison of $A_0(\ptZ)$ for the quark-gluon process for different initial state quarks. Shown are $\Pqu\Pg\to \Pqu\PZ$, $\Pqd\Pg\to \Pqd\PZ$, $\Pqs\Pg\to \Pqs\PZ$, $\Pqc\Pg\to \Pqc\PZ$, and $\Pqb\Pg\to \Pqb\PZ$ quark-gluon processes (with one jet in the final state).

The bottom panel of Fig.~\ref{Fig9} shows a comparison of $A_0(\ptZ)$ for the $\Pqb\Pg\to \Pqb\PZ$ quark-gluon process to \textit{all processes} $\to \Pqb\PZ$ (with one jet in the final state). This indicates that the validity of the \qglu MC prediction can be investigated experimentally in a sample of events with a single $\Pqb$-tagged jet in the final state, as such events are produced via the \qglu process.

\section{Summary}
We investigate the theoretical predictions, obtained with the {\POWHEG}+{\minnlops} event generator, for the angular distributions of
$\pp \to \PZ \to \ell^+ \ell^-$ events produced via quark--antiquark (\qqbar), quark--gluon (\qglu), and gluon--gluon (\gluglu) initiated
sub-processes at $\sqrt{s}=13\TeV$. We find
\begin{itemize}
\item In proton-proton collisions at  13\TeV, \PZ boson production is dominated by gluon-initiated processes, and the \qqbar channel
      accounts for only about 40\% of the total cross section in the \PZ-mass region.
\item The angular coefficients are very similar (within uncertainties) in the two  rapidity regions $|\yZ|<1$ and $1<|\yZ|<2.4$ for all the sub-processes ($A_0^{\qqbar}$ and $A_2^{\qqbar}$, $A_0^{\qglu}$ and $A_2^{\qglu}$, and $A_0^{\gluglu}$ and
      $A_2^{\gluglu}$). 
\item The violation of the Lam-Tung relation ($A_0 = A_2$) originates from events with more than one jet in the final state, for which $A_2$ is smaller than $A_0$ as expected from additional QCD radiation. This can be verified experimentally by comparing inclusive samples with samples restricted to events with $\le$ 1 jet. 
\item For the \qqbar channel, at low and moderate \ptZ, the extracted coefficient $A_0^{\qqbar}(\ptZ)$ is in good agreement with the geometric
      prediction 
      \begin{equation}
      \nonumber
A_0^{\qqbar} = \frac{\ptsq}{M^2+\ptsq},
\end{equation}
\item The approximation 
\begin{equation}
\nonumber
A_0^{\qglu}  = \frac{5\ptsq}{M^2+5\ptsq},
\end{equation}
      proposed in early theoretical studies, significantly overestimates $A_0^{\qglu}$. This can be tested
      experimentally by measuring the angular coefficients in events with a single $b$-tagged jet in the final state,
      which are predominantly produced via the \qglu process.
      
  \item     The same functional form    also overestimates $A_0^{\gluglu}$.
\item We provide parametrizations of the {\POWHEG}+{\minnlops} predictions for $A_0 (\ptZ)$, $A_0^{\qglu}(\ptZ)$ and $A_0^{\gluglu}(\ptZ)$.
\end{itemize}

\section{Acknowledgments }
Research supported by the U.S. Department of Energy under University of Rochester grant number DE-SC0008475 and by MSIP and NRF (Republic of Korea)

\appendix

\section{The method of moments (for full phase space generator level events)}\label{appendix:moments}
One way to extract the theory predictions for the angular coefficients from the MC sample
is by using the method of moments~\cite{Mirkes:1994eb,Mirkes1}.
In this method the \textit{moment} of a function $m_k(\theta,\phi)$ is defined as 
\begin{equation}
M_k= \langle m_k \rangle=\frac{\iint d\sigma(\pt,y,\theta,\phi)m_k(\theta,\phi)d\cos{\theta}d\phi}
{\iint d\sigma(\pt,y,\theta,\phi)d\cos{\theta}d\phi}.
\label{mom}
\end{equation}

Where, 
\begin{equation}
\label{big}
\begin{aligned}
M_0&\equiv \langle \frac{1}{2}(1-3\cos^2{\theta})\rangle =\frac{3}{20}\left (A_0-\frac{2}{3}\right ) \\
M_1&\equiv \langle \sin{2\theta}\cos{\phi}\rangle =\frac{1}{5}A_1  \\
M_2&\equiv \langle \sin^2{\theta}\cos{2\phi}\rangle =\frac{1}{10}A_2  \\
M_3&\equiv \langle \sin{\theta}\cos{\phi}\rangle =\frac{1}{4}A_3  \\
M_4&\equiv \langle \cos{\theta}\rangle =\frac{1}{4}A_4  \\
M_5&\equiv \langle \sin^2{\theta}\sin{2\phi}\rangle =\frac{1}{5}A_5 \\
M_6&\equiv \langle \sin{2\theta}\sin{\phi}\rangle =\frac{1}{5}A_6   \\
M_7&\equiv \langle \sin{\theta}\sin{\phi}\rangle =\frac{1}{4}A_7 
\end{aligned}
\end{equation}
For a set of simulated events, the integrals of Eq.~\ref{mom} are substituted by sums and the
cross section expression is replaced by a sum over all events.

\begin{equation}
\begin{aligned}
M_k
&=
\frac{\sum_i w_i m_k(\theta_i,\phi_i)}
     {\sum_i w_i}, \\[6pt]
\sigma^2(M_k)
&=
\frac{1}{\left(\sum_i w_i\right)^2}
\sum_i w_i^2
\left[
m_k(\theta_i,\phi_i)
-
M_k
\right]^2 .
\end{aligned}
\label{mom2}
\end{equation}  

where $w_i$ is the event weight and $\sigma(M_k)$ is the uncertainty in $M_k$. The set of equations in Eq.~\ref{big} are then solved for the angular coefficients (substituting the moments
with the discrete expressions from Eq.~\ref{mom2}).

\section{Other notations}
Some papers use an older notation\cite{Lam:1978pu,Chang:2018pvk}:
\begin{eqnarray}
\frac{d{\sigma}}{d\cos{\theta}}  \propto  \left (1+\frac{A_0}{2}\right )\left [  1+ \alpha_2 \cos^2{\theta} +\alpha_1\cos{\theta}\right ]
\end{eqnarray}
where
\begin{eqnarray}
\alpha_2 =  \lambda =  \frac {2-3A_0} {2+A_0},~~~A_0 =  \frac {2(1-\lambda)} {3+\lambda}, ~~~\alpha_1 =   \frac {2A_4} {2+A_0} \nonumber
\end{eqnarray}
The following notation is also used: 
\begin{eqnarray}
\frac{d{\sigma}}{d\cos{\theta}d\phi} & \propto &  \left (\frac{1}{\lambda+3}\right ) [1+\lambda \cos^2{\theta} \\ \nonumber
&&+ \mu~ \sin{2\theta}\cos{\phi} +  \frac{\nu}{2}\sin^2{\theta}\cos{2\phi}]
\end{eqnarray}
\begin{eqnarray}
\lambda =  \frac {2-3A_0} {2+A_0},~~~\mu =  \frac {2A_1} {2+A_0}, ~~~\nu =   \frac {2A_2} {2+A_0} \nonumber
\end{eqnarray}
The leading-order L-T relation $A_0 = A_2$ is equivalent to $1 -\lambda -2\nu =0$, and 
$A_0-A_2$ is one of the invariant quantities~\cite{peng2} which is independent of frame.

\section{QCD Scale Uncertainty}

The QCD scale uncertainties on the angular coefficients \(A_0\) and \(A_2\) are evaluated in the rapidity region \(|y|<2.4\), including events with any jet multiplicity and all the sub-processes.  The renormalization and factorization scales, $\mu_R$ and $\mu_F$, are each
varied independently by a factor of 2, up and down, such that their ratio remains within $0.5 <\mu_R/\mu_F< 2.0$. The maximum deviation among these six variants relative to the nominal choice is assigned as a systematic uncertainty associated with the missing higher-order QCD corrections  in the {\POWHEG}+{\minnlops} predictions. These scale uncertainties (which are given in Tables~\ref{tab:A0_scale} and~\ref{tab:A2_scale}) are not included in the uncertainties shown in any of the plots.

\begin{table}[htbp]
\centering
\begin{tabular}{c c c c c}
\hline
\(\ptZ\) [GeV] & \(A_0\) & stat. err. & \(\Delta_{\text{up}}\) & \(\Delta_{\text{down}}\) \\
\hline
0--10   & 0.018396 & 0.000181 & 0.002046 & 0.001251 \\
10--20  & 0.074214 & 0.000222 & 0.005440 & 0.006211 \\
20--35  & 0.190131 & 0.000262 & 0.009704 & 0.011570 \\
35--55  & 0.366987 & 0.000345 & 0.016961 & 0.017386 \\
55--80  & 0.549594 & 0.000482 & 0.031771 & 0.025525 \\
80--120 & 0.705765 & 0.000653 & 0.035724 & 0.025747 \\
120--200& 0.841197 & 0.001004 & 0.019855 & 0.016593 \\
200--400& 0.936367 & 0.002114 & 0.008278 & 0.007367 \\
400--500& 0.988597 & 0.010111 & 0.006561 & 0.005269 \\
500--600& 0.953607 & 0.017803 & 0.003638 & 0.003218 \\
\hline
\end{tabular}
\caption{QCD scale uncertainty for the angular coefficient \(A_0\).}
\label{tab:A0_scale}
\end{table}

\begin{table}[htbp]
\centering
\begin{tabular}{c c c c c}
\hline
\(\ptZ\) [GeV] & \(A_2\) & stat. err. & \(\Delta_{\text{up}}\) & \(\Delta_{\text{down}}\) \\
\hline
0--10   & 0.000814 & 0.000285 & 0.000904 & 0.001720 \\
10--20  & 0.036023 & 0.000351 & 0.003600 & 0.007223 \\
20--35  & 0.124448 & 0.000422 & 0.008013 & 0.012111 \\
35--55  & 0.280448 & 0.000570 & 0.010014 & 0.015068 \\
55--80  & 0.456057 & 0.000819 & 0.005859 & 0.015593 \\
80--120 & 0.617881 & 0.001139 & 0.001587 & 0.016913 \\
120--200& 0.753013 & 0.001798 & 0.007199 & 0.017393 \\
200--400& 0.853672 & 0.003857 & 0.012829 & 0.016917 \\
400--500& 0.894803 & 0.018725 & 0.015463 & 0.020577 \\
500--600& 0.856281 & 0.032712 & 0.016811 & 0.024451 \\
\hline
\end{tabular}
\caption{QCD scale uncertainty for the angular coefficient \(A_2\).}
\label{tab:A2_scale}
\end{table}

\bibliographystyle{apsrev4-1}
\bibliography{Angular_PRD}

\end{document}